\documentclass[fleqn,usenatbib]{mnras}

\usepackage{newtxtext,newtxmath}

\usepackage[T1]{fontenc}

\DeclareRobustCommand{\VAN}[3]{#2}
\let\VANthebibliography\thebibliography
\def\thebibliography{\DeclareRobustCommand{\VAN}[3]{##3}\VANthebibliography}

\usepackage{graphicx}	
\usepackage{amsmath}	

\title[Photodesorption of SO$_2$ ices]{Photodesorption of S-bearing ices I: SO$_2$}
\title[Photodesorption of SO$_2$ ices]{Photodesorption of SO$_2$ and SO from UV-irradiated SO$_2$ ices}

\author[R. Mart\'in-Dom\'enech et al.]{
Rafael Mart\'in-Dom\'enech,$^{1}$\thanks{E-mail: rmartin@cab.inta-csic.es}
Bruno Escribano,$^{1}$
David Navarro-Alamida,$^{1}$
Angèle Taillard,$^{1}$
\newauthor H\'ector Carrascosa,$^{1}$
Guillermo M. Mu\~noz Caro,$^{1}$
Asunci\'on Fuente$^{1}$\\
$^{1}$Centro de Astrobiolog\'ia (CSIC-INTA) 
Carretera de Ajalvir, km. 4, Torrej\'on de Ardoz, E-28850, Madrid, Spain\\
}

\date{Accepted XXX. Received YYY; in original form ZZZ}

\pubyear{\the\year{2025}}

\begin{document}
\label{firstpage}
\pagerange{\pageref{firstpage}--\pageref{lastpage}}
\maketitle

\begin{abstract}
The detection of high gas-phase abundances of SO$_2$ and SO in the cold envelope of an intermediate mass protostar suggests that these molecules might form on icy dust grains and subsequently desorb to the gas phase by non-thermal desorption processes such as photodesorption. 
In this work we report photodesorption yields for SO$_2$ and, tentatively, SO upon ultraviolet photon irradiation of SO$_2$ ice samples at temperatures between 14 and 80 K. 
Photodesorption yields were measured directly in the gas phase using a calibrated quadrupole mass spectrometer. 
Yields of $\sim$2.3 $\times$ 10$^{-4}$ molecule photon$^{-1}$ and $\sim$6 $\times$ 10$^{-5}$ molecule photon$^{-1}$ were estimated for SO$_2$ and SO at 14 K (respectively). 
The SO$_2$ photodesorption yield increased with temperature up to a value of 3.8 $\times$ 10$^{-4}$ molecule photon$^{-1}$ at 70 K, followed by a decrease at 80 K 
that could be due to crystallization of the sample. 
The signal assigned to SO photodesorption did not significantly change with temperature. 
%
The estimated photodesorption yields were included in the \texttt{Nautilus} gas-grain chemical model to evaluate their contribution to the SO$_2$ and SO gas-phase abundances in an astrophysical environment. 
In addition, we also present a theoretically estimated band strength for the $\sim$1395 cm$^{-1}$ SO$_3$ IR feature ($A$ = 1.1 $\times$ 10$^{-16}$ cm molecule$^{-1}$). SO$_3$ is the main detected product in irradiated SO$_2$ ices, and a potential contributor to the $\sim$7.2 $\mu$m band observed in some interstellar ice IR spectra. 
\end{abstract}

\begin{keywords}
astrochemistry,
ISM: abundances, 
ISM: clouds, 
ISM: lines and bands, 
ISM: molecules
\end{keywords}



\section{Introduction}

Sulfur dioxide (SO$_2$) and sulfur monoxide (SO) are two of the most abundant S-bearing molecules in the gas phase of the interstellar medium (ISM) \citep[see, e.g.,][]{charnley97,ferrante08,maity13,elakel22}. 
In particular, these molecules are considered primary tracers for warm chemistry in compact regions around protostars 
\citep[see, e.g.,][]{vandertak03,villarmois19,elakel22}. 
In addition, they are often used to trace shocked regions \citep[see, e.g.,][]{podio15,villarmois18}, including weak accretion shocks in front of the centrifugal barrier in disk-forming protostars \citep[][]{sakai14a,sakai14b,sakai16}.
On the other hand, detections of SO$_2$ and SO in protoplanetary disks have been scarce \citep[][]{fuente10,guilloteau16,booth18,booth21,legal21,booth23}, probably due to the high C/O ratios measured in many disks \citep[][]{dutrey11,semenov18,facchini21}. 
In our Solar System, SO$_2$ and SO have been detected in comets \citep[][]{boissier07,leroy15}, on the surface of Jovian moons 
\citep[see, e.g.,][]{dalton10}, and in the atmosphere of Venus \citep[][]{pollack80,yung82}. 
Due to its prevalence across all stages of the star formation process, understanding the chemical evolution of SO$_2$ and SO is a central question in astrochemistry.  

High gas-phase abundances of SO$_2$ and SO were unexpectedly detected in the cold envelope of the Cyg X-N12 intermediate mass protostar \citep{elakel22}, challenging our current understanding of the interstellar chemistry of these two molecules. 
The high abundances combined with the low diversity of S-O molecules detected in this region suggested that SO$_2$ and SO were formed on icy dust grains and subsequently desorbed to the gas phase through a nonthermal desorption process.  
In this regard, SO$_2$ has been tentatively detected in interstellar ices \citep[][]{boogert97,mcclure23}, while some astrochemical models predict that SO could be the most abundant S-bearing ice molecule \citep[][]{laas19}.
Potential non-thermal desorption processes for these two molecules include chemical desorption, cosmic-ray sputtering, and UV photon-induced desorption. 
However, the dominant contribution cannot be constrained because the non-thermal desorption yields of SO$_2$ and SO are currently unknown \citep[][]{elakel22}. 

Processing of interstellar ices with photons can induce photochemical reactions and photodesorption of both, the initial ice components and the resulting photoproducts. 
Photodesorption is expected to be significant in protoplanetary disks, the outer layers of dense clouds, and photon-dominated regions (PDRs) in general, but even in the interior of dense clouds it can have a comparable contribution to other non-thermal desorption processes \citep[see, e.g.,][]{oberg07,munozcaro10}. 
Therefore, it is fundamental to constrain the photodesorption yields of S-bearing species in order to explain their gas-phase abundances in cold regions of the ISM, such as the cold envelope of Cyg X-N12. 

Photodesorption yields can be measured in the laboratory under conditions relevant to the ISM \citep[see, e.g.,][]{oberg07,munozcaro10,bertin12,chen14,fillion14,martin15,gustavo16,martin18}. 
Thus far, the photodesorption yield of only one S-bearing molecule has been reported in the literature \citep[H$_2$S,][]{fuente17}. 
Photodesorption of SO$_2$ was studied in \citet{burke05} and \citet{wesemberg07}, but the experimental conditions were not completely relevant to the ISM.  
In those experiments, SO$_2$ films were adsorbed on silver and silicon surfaces at 89$-$95 K and irradiated with photons in the 260 $-$ 530 nm range. 
The authors suggested that under those conditions photodesorption was substrate-mediated, i.e., it probably took place upon absorption of photons by the metal surface and subsequent interaction with the SO$_2$ film.  
Other works evaluating the photoprocessing of SO$_2$ ice samples were focused on the induced photochemistry, but did not report any photodesorption measurements \citep[][]{schriver03,deSouza17}. 

In this work we have experimentally measured photodesorption yields for SO$_2$ and, tentatively, SO upon ultraviolet (UV) photon irradiation of SO$_2$ ice samples at different temperatures relevant to the interior of dense clouds. 
In addition, we have theoretically estimated the band strength of the IR feature corresponding to SO$_3$, the main detected product in irradiated SO$_2$ ices. 
Section \ref{sec:methods} presents the experimental setup used for the laboratory measurements, as well as the density functional theory (DFT) calculations used for the theoretical estimation of the band strength. 
The measured photodesorption yields and the calculated IR band strength are presented in Sect. \ref{sec:results} and discussed in Sect. \ref{sec:disc}. Finally, the main conclusions are listed in Sect. \ref{sec:conc}.

\section{Methods}\label{sec:methods}

The experiments carried out for this work are listed in Table \ref{tab:exp}. 
These experiments were performed with the InterStellar Astrochemistry Chamber (ISAC) setup at the Centro de Astrobiolog\'ia (CAB, CSIC-INTA). 
ISAC consists of an ultra-high-vacuum (UHV) chamber with a base pressure on the order of $\sim$5 $\times$ 10$^{-10}$ mbar, similar to that found in dense cloud interiors. 
The ISAC setup is described in detail in \citet{munozcaro10}, and the main features relevant to the experiments presented in this work are highlighted below. 
In summary, a series of SO$_2$ ice samples (Sect. \ref{sec:exp_ice}) were irradiated with UV photons (Sect. \ref{sec:exp_irr}) in order to measure the resulting SO$_2$ and SO photodesorption yields. During irradiation, photodesorbing molecules were detected in the gas phase and quantified using a quadrupole mass spectrometer (QMS, Sect. \ref{sec:exp_qms}). 
The irradiated ice samples were also monitored with infrared (IR) spectroscopy (Sect. \ref{sec:exp_IR}), that allowed us to detect the formation of SO$_3$ molecules. The band strength of the detected SO$_3$ IR feature was theoretically estimated using DFT calculations (Sect. \ref{sec:exp_dft}).

\begin{table*}
    \centering
    \begin{tabular}{ccccccccc}
        & Dep. T & Irrad. T & Initial thickness & \multicolumn{2}{c}{Irrad. Fluence} & k$_{CO}$ & Y$_{pd}$ (SO$_2$) & Y$_{pd}$ (SO)\\
       Exp. & (K) & (K) & (ML) & ($\times$10$^{17}$ ph. cm$^{-2}$) & ($\times$10$^{18}$ eV cm$^{-2}$) & ($\times$10$^{-11}$ A min ML$^{-1}$) & \multicolumn{2}{c}{($\times$10$^{-4}$ mol ph$^{-1}$)} \\
       \hline
       1 & 14 & 14 & 50 & 3.5 & 2.8 & 3.2 & 2.4 & 0.4\\ 
       2 & 14 & 14 & 50 & 3.6 & 2.9 & 4.0 & 2.3 & 0.5\\ 
       3 & 14 & 14 & 50 & 3.6 & 2.9 & 5.8 & 2.3 & 0.8\\ 
       4 & 80 & 13 & 45 & 3.4 & 2.7 & 4.0 & 1.8 & 0.5\\ 
       5 & 81 & 14 & 60 & 3.4 & 2.7 & 6.7 & 2.0 & 0.6\\ 
       6 & 22 & 22 & 50 & 3.5 & 2.8 & 5.7 & 2.1 & 0.5\\ 
       7 & 22 & 22 & 50 & 3.4 & 2.7 & 6.4 & 2.2 & 0.6\\ 
       8 & 31 & 31 & 50 & 0.9 & 0.7 & 5.6 & 2.7 & 0.8\\ 
       9 & 41 & 41 & 50 & 3.6 & 2.9 & 5.3 & 2.9 & 0.6\\ 
       10 & 60 & 60 & 60 & 3.4 & 2.7 & 5.3 & 3.5 & 0.6\\
       11 & 70 & 70 & 50 & 3.5 & 2.8 & 6.1 & 3.8 & 0.8\\
       12 & 81 & 81 & 50 & 3.4 & 2.7 & 4.6 & 3.1 & 0.5\\
       13 & 80 & 80 & 55 & 3.6 & 2.9 & 6.7 & 3.3 & 0.6\\
       14 & 80 & 80 & 50 & 4.1 & 3.3 & 7.4 & 2.9 & 0.5\\
       \hline
    \end{tabular}
    \caption{Summary of the performed irradiation experiments. 
    The ice thickness was measured at the deposition temperature, and presented a 5\% uncertainty (see the text). 
    Note that 1 ML = 10$^{15}$ molecules cm$^{-2}$.
    A 20\% systematic uncertainty was assumed for the $k_{CO}$ proportionality constant (see the text). 
    A 20\% and 40\% uncertainty was assumed for the listed SO$_2$ and SO photodesorption yields (Y$_{pd}$), respectively (see the text).
    }
    \label{tab:exp}
\end{table*}

\subsection{Ice sample preparation}\label{sec:exp_ice}

The SO$_2$ ice samples were grown on a MgF$_2$ substrate located at the center of the UHV chamber. 
Deposition took place by exposing the substrate to SO$_2$ molecules (gas, 99.9\%, Air Liquide) introduced into the chamber from an independently pumped gas line assembly. 
Ice samples were deposited and irradiated at a temperature between 14 and 80 K (second and third columns in Table \ref{tab:exp}) to study the effect of ice temperature and structure on the measured photodesorption yields. 
We note that, according to previous studies, the transition from amorphous to crystalline SO$_2$ ice takes place around 70 K \citep{schriver03a}. 
Therefore, we were able to evaluate the effect of the ice structure independently from the ice temperature by comparing the results obtained in Experiments 4 and 5 (crystalline ices deposited at 80 K and irradiated at 14 K) with those obtained in Experiments 1$-$3 (amorphous ices deposited and irradiated at 14 K). 
The initial thickness of the ice samples was 45$-$60 monolayers (1 ML = 10$^{15}$ molecules cm$^{-2}$), as determined by infrared (IR) spectroscopy (Sect. \ref{sec:exp_IR}) and indicated in the fourth column of Table \ref{tab:exp}. 

\subsection{UV photon irradiation of the ice samples}\label{sec:exp_irr}

\begin{figure}
    \centering
    \includegraphics[width=0.65\linewidth]{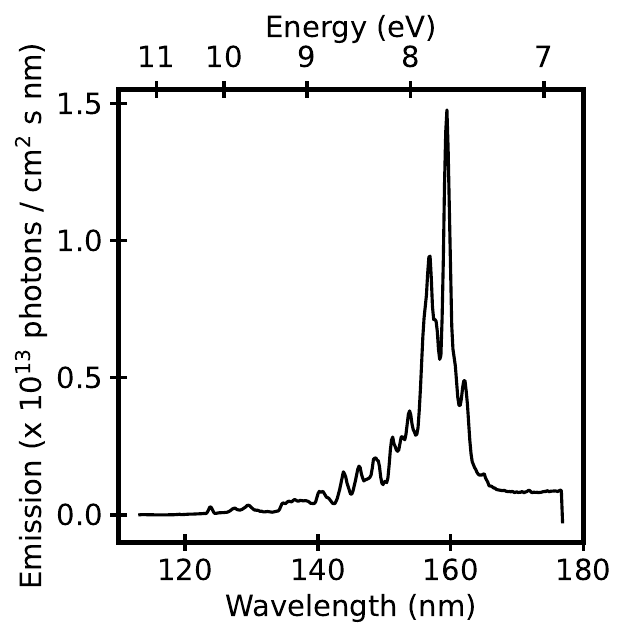}
    \caption{Emission spectrum of the MDHL measured in ISAC.}
    \label{fig:uvspec}
\end{figure}

The deposited ice samples were irradiated using a F-type microwave-discharged hydrogen-flow lamp (MDHL) from Opthos Instruments. 
The vacuum-ultraviolet (VUV) photons from the lamp entered the chamber through a MgF$_2$ window that absorbed all photons with $\lambda$ $<$ 114 nm \citep[][]{chen14}. 
A quartz tube in the interior of the chamber acted as a light guide. 

The emission spectrum of the lamp is shown in Fig. \ref{fig:uvspec}. It was measured \textit{in situ} using a McPherson 0.2 m focal length vacuum-ultraviolet (VUV) monochromator (model234/302) with a photomultiplier tube (PMT) detector located behind another MgF$_2$ window at the rear of the chamber \citep[][]{gustavo14a,gustavo14b,gustavo14c,carlos24}. 
Note that the VUV photons that reached the detector during the emission spectrum measurement intersected by two MgF$_2$ windows, while those reaching the ice samples only intersected by one. 
Therefore, the spectrum of the photons irradiating the ice samples could be slightly different from that shown in Fig. \ref{fig:uvspec}. 
The lamp emission spectrum was similar to the expected spectrum of the secondary UV field produced in the interior of dense clouds by the interaction of cosmic rays with gas-phase H$_2$ molecules \citep[][]{gredel89,cecchi92,shen04}. 
The mean energy of the photons was 8 eV. 
The photon flux was continuously measured during irradiation using a calibrated Ni-mesh located at the end of the quartz tube, and had a value of $\sim$8 $\times$ 10$^{13}$ photons cm$^{-2}$ s$^{-1}$. 
Irradiation was performed in successive intervals of 3$-$30 min, for a total irradiation time of $\sim$70 min. 
This led to an accumulated fluence of $\sim$3.5 $\times$ 10$^{17}$ photons cm$^{-2}$ or $\sim$2.8 $\times$ 10$^{18}$ eV cm$^{-2}$, except for Experiments 8 and 14 with shorter and longer total irradiation times, respectively (see fifth and sixth columns in Table \ref{tab:exp}). 

The UV-absorption cross-section in the 120$-$320 nm range of SO$_2$ ice samples at 25$-$80 K was reported in \citet{holtom06}. 
Cross sections on the order of $\sim$6 $\times$ 10$^{-18}$ cm$^{2}$ were measured in the 7$-$9 eV range (where most of the emission of the MDHL is located, see Fig. \ref{fig:uvspec}) assuming a density of 1.92 g cm$^{-3}$ for the SO$_2$ ice. 
More recently, \citet{yarnall22} reported a value of 1.395 g cm$^{-3}$ for the density of solid SO$_2$ at 19 K. 
Correcting the UV-absorption cross-section reported in \citet{holtom06} with the more recent value of the density leads to slightly higher absorption cross-sections (close to $\sim$8 $\times$ 10$^{-18}$ cm$^{2}$). 
Using the VUV monochromator and PMT mentioned above, we measured the UV-absorption cross-section of a SO$_2$ ice sample and obtained comparable values (not shown). 
As a result, we estimated that roughly $\sim$30\% of the incident photons were absorbed in Experiments 1$-$14.

\subsection{Quadrupole mass spectrometry of the photodesorbing molecules}\label{sec:exp_qms}

\begin{figure}
    \centering
    \includegraphics[width=0.6\linewidth]{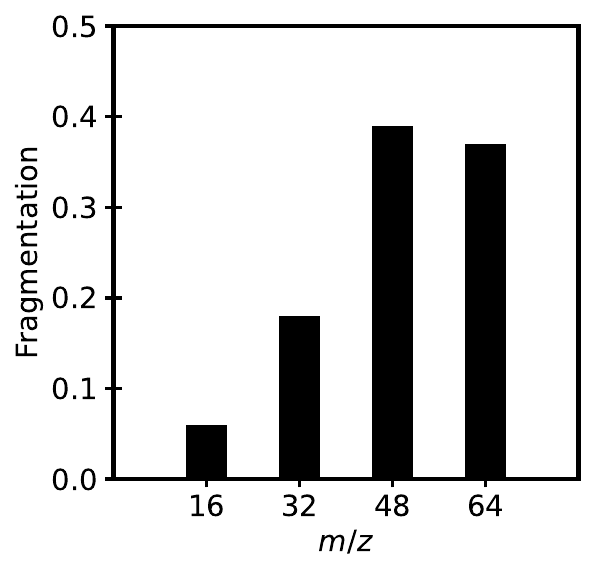}
    \caption{Fragmentation pattern of SO$_2$ molecules measured in ISAC for pressures $\le$ 2 $\times$ 10$^{-8}$ mbar.}
    \label{fig:so2_MS}
\end{figure}

\begin{table}
    \centering
    \begin{tabular}{ccccc}
    Factor & CO & \multicolumn{2}{c}{SO$_2$} & SO\\
        \hline
    monitored $m/z$ & 28 & 64 & 48 & 48\\
    $\sigma^+(X)$ (\AA$^2$)$^a$ & 2.516 & \multicolumn{2}{c}{4.992} & 4.992\\
    $F_F(m)^b$ & 0.90 & 0.37 & 0.39 & 0.65\\
    $S(28)/S(m/z)^c$ & 1.00 & 1.24 & 1.06 & 1.06\\
    $S_{rel}(X)^d$ & 1.00 & \multicolumn{2}{c}{0.67} & 0.82\\ 
    \hline
    \end{tabular}
    \caption{Parameters used in Eq. \ref{eqn_qms} to estimate photodesorbing column densities from measured QMS signals. 
    $^a$Extracted from NIST database, except for SO, for which the SO$_2$ value was adopted as a first approximation \citep{dols24}.
    $^b$Measured in ISAC, except for SO, extracted from \citet{dols24}. 
    $^c$This work (Appendix \ref{sec:appA}).
    $^d$Calculated with Eq. \ref{eq_srel}.
    }
    \label{tab:qms_param}
\end{table}

During irradiation, the molecules photodesorbing from the ice samples to the gas phase were detected with a Pfeiffer Prisma QMS. 
Molecules reaching the QMS were ionized by $\sim$70 eV electron bombardment, which led to fragmentation of a fraction of the detected molecules. 
Therefore, photodesorbing molecules could be identified through their molecular ion (with a mass-to-charge ratio $m/z$ equal to the molecular mass) or through one of their fragments (with a lower $m/z$).  
Figure \ref{fig:so2_MS} shows the fragmentation pattern (also known as mass spectrum) of SO$_2$ molecules measured in ISAC for pressures $\le$ 2 $\times$ 10$^{-8}$ mbar (at higher pressures the molecular ion $m/z$ = 64 presented a higher intensity than the $m/z$ = 48 fragment, similar to the mass spectrum reported in the NIST database). 

In order to quantify the photodesorption yields, the signal measured by the QMS for a particular molecular ion or fragment during irradiation ($I(m/z)$) was converted into a photodesorbing column density ($N(X)$) using Eq. \ref{eqn_qms} \citep{martin15,martin16}: 

\begin{equation} 
N(X) = \frac{I(m/z)}{k_{CO}} \cdot \frac{\sigma^+(CO)}{\sigma^+(X)} 
\cdot \frac{F_F(28)}{F_F(m)} \cdot \frac{S(28)}{S(m/z)} \cdot S_{rel}(X),  \label{eqn_qms}
\end{equation}

where 
$k_{CO}$ was a proportionality constant measured for CO molecules, 
$\sigma^+(X)$ was the electron-impact ionization cross-section for species $X$, 
$F_F(m)$ was the fraction of molecules $X$ leading to a fragment of mass $m$ in the QMS,  
$S(m/z)$ was the sensitivity of the QMS to a fragment with mass-to-charge ratio $m/z$, 
and $S_{rel}(X)$ was the relative pumping speed of molecule $X$ with respect to CO. 
The relative pumping speed was calculated as

\begin{equation}
    S_{rel}(X) = 1.258 - 9.2 \times 10^{-3} \times M(X),  \label{eq_srel}
\end{equation}

\noindent (where M(X) was the molecular mass of molecule $X$) according to the manufacturer of the pumping devices used in the ISAC setup \citep[see][for more information]{kaiser95,martin16}. 
The $k_{CO}$ and $S(m/z)$ parameters were derived from dedicated calibration experiments presented in Appendix \ref{sec:appA}. 
While carrying out the experiments listed in Table \ref{tab:exp}, we noticed a change in the relative $k_{CO}$ proportionality constant from one experiment to another, independent from the estimated 20\% systematic uncertainty (Appendix \ref{sec:appA}). 
Instead of running a different calibration experiment every day, we applied a correction factor to the value measured from the calibration experiment described in Appendix \ref{sec:appA} to obtain the values corresponding to Experiments 1$-$14 (seventh column of Table \ref{tab:exp}). 
The rest of parameters required in Eq. \ref{eqn_qms} are listed in Table \ref{tab:qms_param}.

If the measured ion current in ampers was used as $I(m/z)$ in Eq. \ref{eqn_qms}, the calculated $N(X)$ was the photodesorbing column density per unit time, i.e., the photodesorption rate. 
In order to estimate the photodesorption yields, we integrated the measured ion currents over the irradiation time for every irradiation interval using the \texttt{integrate.simps} function in the SciPy library, and used Eq. \ref{eqn_qms} to obtain the photodesorbed column density in every interval. We then divided this column density by the irradiated fluence in the corresponding interval to calculate the number of photodesorbed molecules per incident photon. The photodesorption yields listed in Table \ref{tab:exp} correspond to the mean values for all irradiation intervals for each experiment. 
In addition to the systematic 20\% uncertainty estimated for $k_{CO}$, we assumed an additional 6\% and 35\% experimental uncertainty that accounted for the differences in the photodesorption yields calculated for SO$_2$ and SO (respectively) in Experiments 1$-$3 and 12$-$14 (that were carried out under the same conditions, and were thus expected to present the same results). 
All other sources of uncertainty were considered negligible. 
This led to a total uncertainty of 20\% and 40\% for the SO$_2$ and SO photodesorption yields, respectively. 
Note that when comparing photodesorption yields measured in different experiments, only the experimental uncertainty needed to be considered, since the systematic uncertainty in $k_{CO}$ was the same for all experiments.

\subsection{Infrared ice spectroscopy}\label{sec:exp_IR}

\begin{table*}
    \centering
    \begin{tabular}{ccccc}
    &Wavenumber&Wavelength&Band strength&\\
      Molecule & (cm$^{-1}$) & ($\mu$m) & (cm molecule$^{-1}$) & Reference\\
     \hline 
    SO$_2$ & 1320 & 7.58 & 4.2 $\times$10$^{-17}$ & \citet{yarnall22}\\ 
    SO$_2$ & 1150 & 8.70 & 7.3 $\times$10$^{-18}$ & \citet{yarnall22}\\ 
    SO$_3$ & 1395 & 7.17 & 1.1 $\times$10$^{-16}$ & This work\\ 
    \end{tabular}
    \caption{Band strengths of selected features in pure ice IR spectra. Uncertainties of 5\% and 15\% were assumed for the SO$_2$ and SO$_3$ IR band strengths, respectively (see the text). 
    }
    \label{tab:ir}
\end{table*}

Ice samples were also monitored through IR spectroscopy in transmittance using a Vertex 70 Bruker Fourier Transform IR (FTIR) spectrometer with a DTGS detector. The spectra were averaged over 128 interferograms and collected with a resolution of 2 cm$^{-1}$ in the 5000$-$1000 cm$^{-1}$ range. 
Equation  \ref{eqn} was used to calculate the ice column density of SO$_2$ and the detected photoproduct SO$_3$ 
using the integrated absorbance of the corresponding IR bands: 

\begin{equation}
N=\frac{1}{A}\int_{band}{\tau_{\nu} \ d\nu},
\label{eqn}
\end{equation}

\noindent where $N$ is the column density in molecules cm$^{-2}$,  $\tau_{\nu}$ is the integrated optical depth of the absorption band (2.3 times the absorbance), and $A$ is the band strength of the IR feature in cm molecule$^{-1}$.
IR bands were numerically integrated using the \texttt{integrate.simps} function in the SciPy library.
Band strengths of the SO$_2$ IR features are reported in the literature \citep{yarnall22}. 
However, an IR band strength for SO$_3$ 
had not been previously reported, and the value for the SO$_2$ feature at 1323 cm$^{-1}$ was usually assumed in previous works \citep[see, e.g.,][]{garozzo08,deSouza17}. 
As part of this work we theoretically calculated the SO$_3$ IR band strength (Sect. \ref{sec:exp_dft}). 
Band strengths values are listed in Table \ref{tab:ir}. 
Note that using the SO$_2$ band strength to calculate the SO$_3$ ice column density would overestimate the latter by at least a factor of two. 
Uncertainties in ice column densities derived from Eq. \ref{eqn} are dominated by band strength uncertainties, as the contribution from the integrated absorbance uncertainty is considered negligible. 
\citet{yarnall22} estimated an uncertainty of 5\% for the SO$_2$ IR band strengths, while a 15\% uncertainty was assumed for SO$_3$ (Sect. \ref{sec:exp_dft}).

\subsection{Theoretical estimation of the SO$_3$ IR band strength}
\label{sec:exp_dft}
The band strength of the $\sim$1395 cm$^{-1}$ SO$_3$ IR feature was obtained from DFT calculations \citep{hohenberg64,kohn65}. 
To this purpose, an amorphous SO$_3$ ice was first modeled in a simulation box.
The amorphous ice model was comprised of eight SO$_3$ molecules with randomized initial positions in a cubic box. We applied periodic boundary conditions to simulate an infinite amorphous ice. Some simulations were tested with 16 molecules per box but showed negligible differences with the smaller system, while being computationally much more expensive.
An ice temperature of 10 K was subsequently simulated by applying 100 ps of NVT molecular dynamics (MD) before the DFT geometry optimization, allowing the molecules to move and interact through central potentials modeling the Coulomb and van der Waals forces.
Geometry optimization was then performed with the \texttt{CASTEP} software \citep{CASTEP}, using the Perdew-Burke-Ernzerhof (PBE) version of the Generalized Gradient Approximation \citep[GGA,][]{PBE}. 
Long range dispersion correction was implemented using the Grimme method \citep{SEDC-G06}. 
For these calculations, the energy convergence parameter was set to 10$^{-6}$ eV/{\AA} and the cut-off was set to 925 eV.
The resulting system is shown in Fig. \ref{fig:SO3_box}
After applying the MD trajectory at 10 K and the geometry optimization with DFT, IR vibrations were computed using linear perturbation theory \citep{CASTEP_DFPT} as implemented in \texttt{CASTEP}. 
This method provides a series of discrete, adjacent IR absorption bands corresponding to similar vibrations of the molecules contained in the simulation box. 
For every discrete band, the absorption intensity is measured in cm molecule$^{-1}$. 
The corresponding $\sim$1395 cm$^{-1}$ band strength was calculated by adding the IR absorption intensity of all discrete vibrations corresponding to the anti-symmetric stretching mode, and dividing by the number of molecules in the simulation box. 
This method provides overall good agreement 
with experimental results for many amorphous ices, both for IR band positions and absorbance intensities \citep[\textit{in prep.}]{bruno25}.

A critical parameter of this method is the density of the simulation box, that affects the calculated IR vibrations and their band strengths. 
Unfortunately, the density of amorphous SO$_3$ ice is currently unknown, but it is expected to be close to the density of the liquid phase (as it is the case of other amorphous ices, see Appendix \ref{sec:appDFT}). 
Therefore, we assumed the reported density for liquid SO$_3$ in the International Labour Organization (ILO) database (1.92 g cm$^{-3}$) as the density of an amorphous SO$_3$ ice. 
The uncertainty in the derived IR band strength related to the assumption of the SO$_3$ ice density was estimated to be 15\%. This is explained with more detail in Appendix \ref{sec:appDFT}. The value of the band strength is presented in Table \ref{tab:ir}. 

\section{Results}\label{sec:results}

\begin{figure*}
    \centering
    \includegraphics[width=0.7\linewidth]{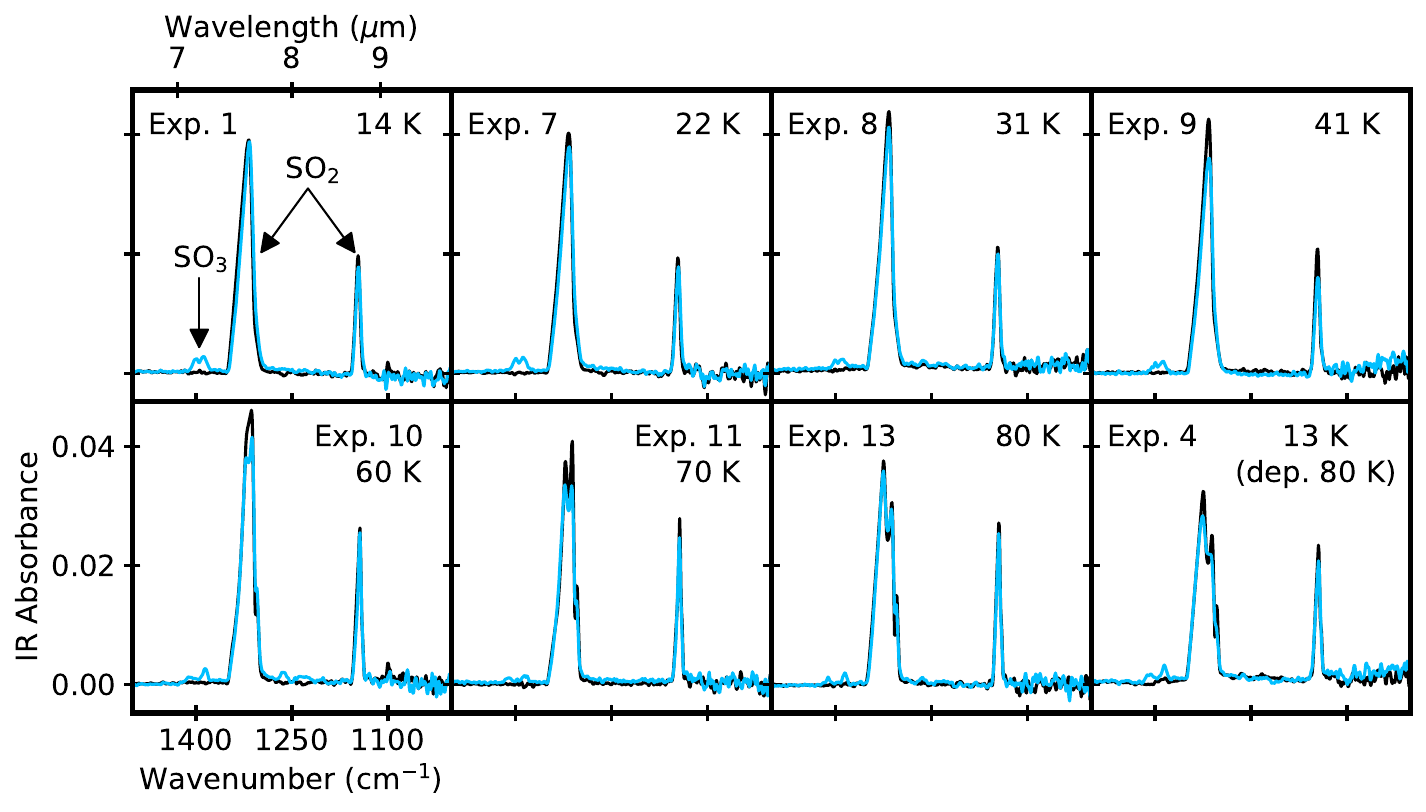}
    \caption{IR spectra in the 1500$-$1000 cm$^{-1}$ range of SO$_2$ ice samples deposited at 14$-$80 K before (black) and after (blue) VUV photon irradiation. IR bands corresponding to SO$_2$ and SO$_3$ are indicated in the first panel.}
    \label{fig:so2_IR}
\end{figure*}

\begin{figure*}
    \centering
    \includegraphics[width=0.7\linewidth]{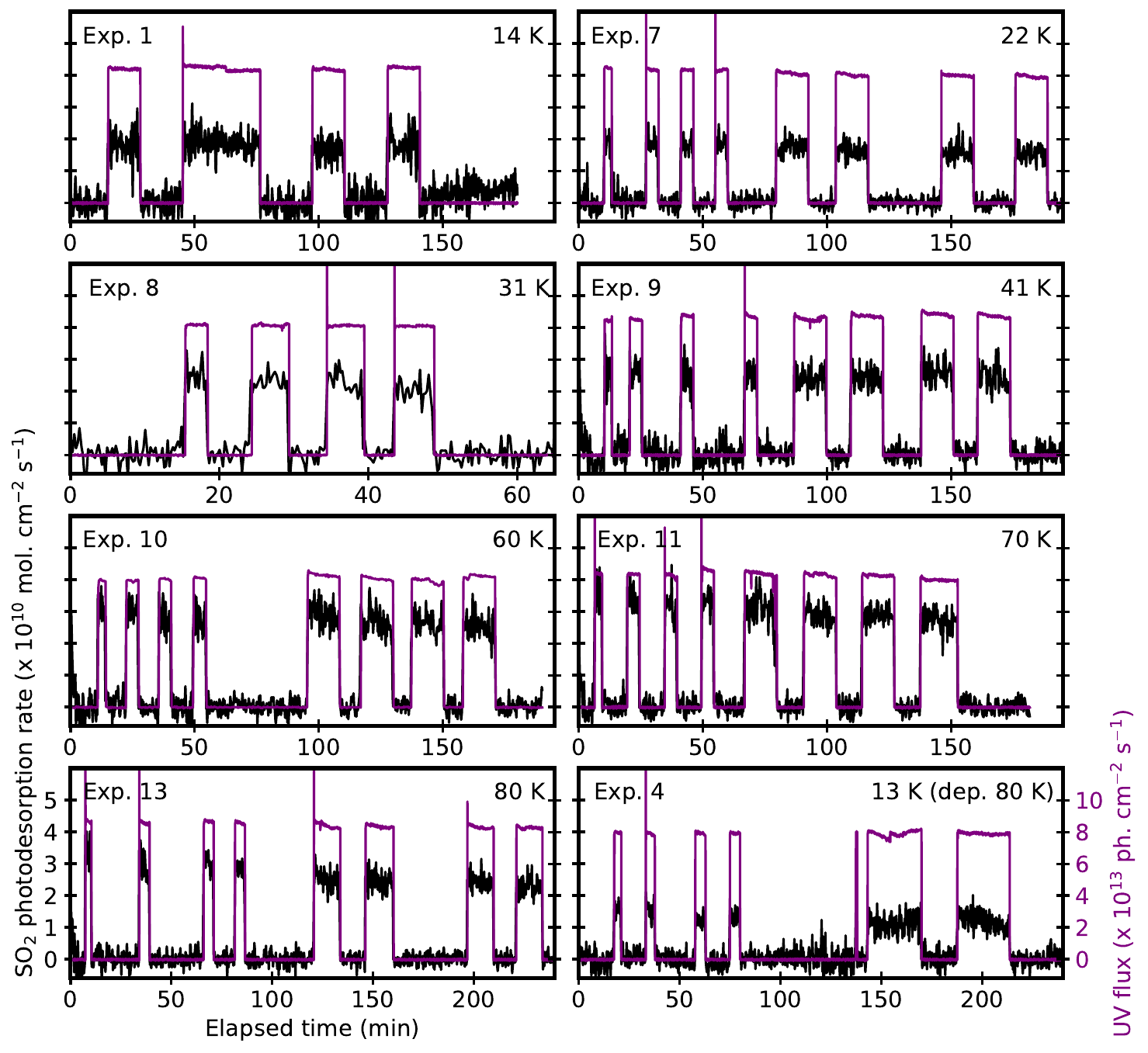}
    \caption{SO$_2$ photodesorption rate (calculated from the baseline-subtracted $m/z$ = 64 ion current using Eq. \ref{eqn_qms}) during irradiation of SO$_2$ ice samples at 13$-$80 K (black), along with the irradiated UV flux in every experiment (purple).}
    \label{fig:so2_QMS_bl}
\end{figure*}

\begin{figure*}
    \centering
    \includegraphics[width=0.7\linewidth]{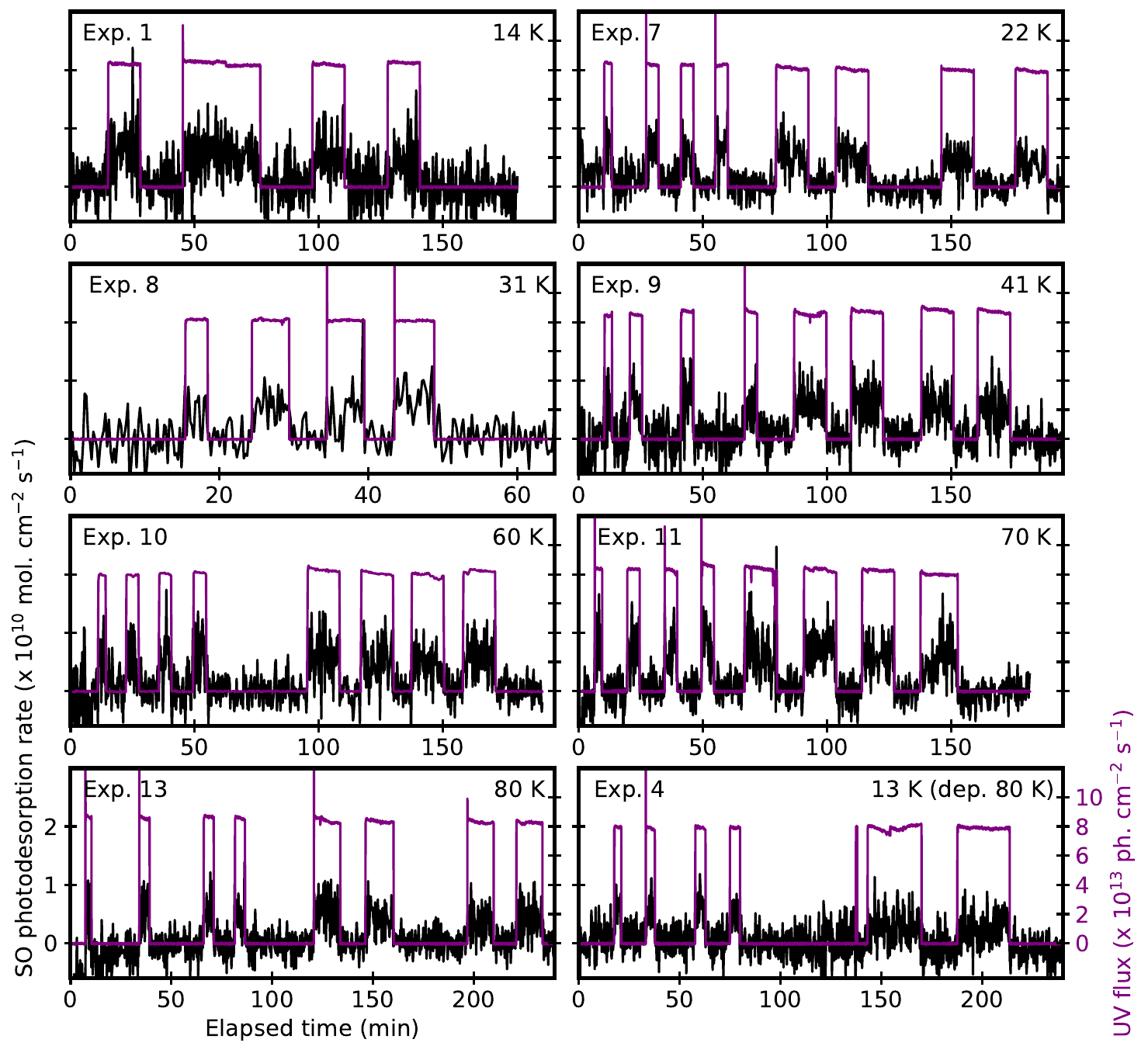}
    \caption{Photodesorption rate assigned to SO (calculated from the measured $m/z$ = 48 ion current using Eq. \ref{eqn} after subtracting the contribution from the SO$_2$ fragmentation) during irradiation of SO$_2$ ice samples at 13$-$80 K (black), along with the irradiated UV flux in every experiment (purple).}
    \label{fig:so_QMS_corr}
\end{figure*}

\subsection{IR spectroscopy of the irradiated SO$_2$ ice samples}\label{sec:results_IR}
The IR spectra in the 1500$-$1000 cm$^{-1}$ range of SO$_2$ ice samples deposited at 14$-$80 K are shown in Fig. \ref{fig:so2_IR}. 
Before irradiation, two IR bands were detected at 1320 and 1150 cm$^{-1}$, corresponding to the $\nu_3$ (asymmetric stretch) and $\nu_1$ (symmetric stretch) vibrational modes, respectively. 
The profile of the $\nu_3$ band was related to the ice structure. Amorphous ices presented a single peak, while crystallization was characterized by the splitting of the band into three different components \citep{schriver03a}.   
After irradiation, a decrease of up to 10\% was observed for the absorbance of the SO$_2$ IR bands, due to the combination of ice photodissociation and photodesorption. 
At the same time, we observed the appearance of a double-peaked IR band centered at $\sim$1395 cm$^{-1}$ due to the formation of SO$_3$ molecules \citep[see, e.g.,][for the assginment of the IR feature]{chaabouni00}. 
According to the theoretically calculated IR band strength (Table \ref{tab:ir}), up to 1 ML of SO$_3$ was formed in Experiments 1$-$14. 
Previous works presenting the energetic processing of SO$_2$ ice samples with different sources (VUV photons, X-ray photons, keV electrons, and keV-MeV H$^+$ and He$^+$) also reported the formation of SO$_3$ \citep{moore84,schriver03,moore07,garozzo08,deSouza17,mifsud22}. 
In some cases, formation of O$_3$ was also observed \citep{schriver03,garozzo08}. 
However, the corresponding IR band at $\sim$1040 cm$^{-1}$ was not detected in Fig. \ref{fig:so2_IR}. 
\citet {schriver03} and \citet{garozzo08} were not able to pinpoint the particular conditions required for the formation of O$_3$. \citet{garozzo08} suggested that O$_3$ molecules are only formed after high energy fluences. In this regard, the irradiated fluence in our experiments ($\sim$2.8 $\times$ 10$^{18}$ eV/cm$^2$) was even lower than that reported by \citet{deSouza17} ($\sim$6.0 $\times$ 10$^{20}$ eV/cm$^2$) and \citet{mifsud22} ($\sim$1.2 $\times$ 10$^{20}$ eV/cm$^2$), where O$_3$ was not detected. 
Ice temperature could also play a role, since \citet{garozzo08}  observed O$_3$ formation at 16 K, but not at 80 K (for the same irradiated energy fluence). 
In addition to SO$_3$, formation of SO was also expected in our experiments. However, the corresponding IR feature might overlap with that of SO$_2$, hindering its detection in the IR spectra \citep{schriver03}. 
%

\subsection{Photodesorption yields of SO$_2$ and SO}
Figure \ref{fig:so2_QMS_bl} shows the SO$_2$ photodesorption rate  during irradiation of SO$_2$ ice samples at 13$-$80 K. 
The photodesorption rate was calculated using Eq. \ref{eqn_qms} from the measured $m/z$ = 64 ion current after baseline subtraction.
(the measured $m/z$ = 64 QMS signal without baseline subtraction is shown in Fig. \ref{fig:so2_QMS}). 
Photodesorption of SO$_2$ did not significantly change with the irradiated fluence. 
The corresponding SO$_2$ photodesorption yields (calculated from the integrated $m/z$ = 64 ion current using Eq. \ref{eqn_qms}, and the irradiated fluence) are listed in the eighth column of Table \ref{tab:exp}. 
A photodesorption yield of $\sim$2.3 $\times$ 10$^{-4}$ mol. ph.$^{-1}$ was measured for the ice samples deposited and irradiated at 14 K (Experiments 1$-$3 in Table \ref{tab:exp}).  
The photodesorption yield increased with temperature between 22 and 70 K (Experiments 6$-$11), up to a value of 3.8 $\times$ 10$^{-4}$ mol. ph.$^{-1}$ at 70 K (Exp. 11). This is the temperature in which the transition from amorphous to crystalline structure is expected to take place \citep{schriver03a}. 
In fact, the ice deposited at 70 K seemed to be at least partially crystalline according to the IR spectra shown in Fig. \ref{fig:so2_IR}.  
Finally, the photodesorption yield decreased down to $\sim$3.1 $\times$ 10$^{-4}$ mol. ph.$^{-1}$ at 80 K (Experiments 12$-$14). At this temperature the ice was expected to be completely crystalline.  

In order to evaluate the effect of the ice structure indepentently from the temperature, we deposited crystalline SO$_2$ ice samples at 80 K and irradiated them at 14 K (Experiments 4 and 5). The corresponding SO$_2$ photodesorption yield was slightly lower ($\sim$1.9 $\times$ 10$^{-4}$ mol. ph.$^{-1}$) than that measured for amorphous ice samples deposited and irradiated at 14 K ($\sim$2.3 $\times$ 10$^{-4}$ mol. ph.$^{-1}$, Experiments 1$-$3). 
Therefore, although the effect of the ice temperature on the photodesorption yield appeared to be dominant, there seemed to be also a contribution from the structure of the ice, with crystalline ices presenting slightly lower yields than amorphous ices when irradiated at the same temperature. 
This would explain the observed decrease in the photodesorption yield from 70 K to 80 K. 

Photodesorption of SO$_2$ was also detected via the $m/z$ = 48 QMS signal (Fig. \ref{fig:so_QMS}), due to fragmentation of SO$_2$ molecules into SO$^+$ ions upon arrival to the filament of the QMS (Sect. \ref{sec:exp_qms}). 
However, the measured $m/z$ = 48 ion current was higher than expected based on the SO$_2$ fragmentation pattern shown in Fig. \ref{fig:so2_MS}, and the QMS sensitivity to the $m/z$ = 48 and $m/z$ = 64 ions listed in Table \ref{tab:qms_param}. 
The excess in the measured $m/z$ = 48 QMS signal could be due to additional photodesorption of SO and/or O$_3$ molecules, both potential photoproducts with a molecular mass of 48 amu. 
As explained in Sect. \ref{sec:results_IR}, no IR feature corresponding to O$_3$ was detected in the IR spectra shown in Fig. \ref{fig:so2_IR}. 
Formation of SO was expected, but it could not be confirmed either with the IR spectra, probably due to the overlapping of the corresponding IR band with the $\nu_3$ band of SO$_2$. (Sect. \ref{sec:results_IR}). 
Similar to SO$_2$, photodesorption of SO and/or O$_3$ would have led to an increase of not only the $m/z$ = 48 QMS signal, but also other fragments following their respective fragmentation patterns. 
The O$_3$ mass spectrum was reported in \citet{herron56}, indicating that the $m/z$ = 32 QMS signal should have been $\sim$5 times higher than the excess in the $m/z$ = 48 signal if the latter was due to O$_3$ photodesorption. 
The SO mass spectrum is not reported in the literature, but a fragmentation pattern could be approximated from the SO electron-impact ionization and dissociative-ionization cross sections reported in \citet{dols24}.  
In this case, the $m/z$ = 32 QMS signal should have been more than 2 times lower than the excess in the $m/z$ = 48 signal. 
In our experiments, no significant increase above the baseline was observed for the $m/z$ = 32 QMS signal during irradiation of the ice samples, which was compatible with photodesorption of SO.
Moreover, we note that formation of SO would only require photodissociation of SO$_2$ molecules into SO + O \citep[with an associated dissociation energy of 0.78 eV,][]{mozo07}, whereas formation of O$_3$ would additionally require recombination of three O atoms, which might be a less likely process. 
Therefore, we tentatively assigned the excess in the measured $m/z$ = 48 QMS signal to the photodesorption of SO molecules. 
Experiments with an SO$_2$ isotopolog would be required in order to confirm this assignment. 
Figure \ref{fig:so_QMS_corr} shows the estimated SO photodesorption rate during irradiation of SO$_2$ ice samples at 13$-$80 K, that did not present significant changes with the irradiated fluence (as it was the case of SO$_2$). 
The corresponding photodesorption yields are listed in the ninth column of Table \ref{tab:exp}. 
In this case, no clear trend with temperature was observed. The 4$-$8 $\times$ 10$^{-5}$ mol. ph.$^{-1}$ yields estimated in Experiments 1$-$14 were all similar within the 35\% experimental uncertainty.  
We note that the QMS signal of $m/z$ = 80 corresponding to the molecular ion of SO$_3$ was also monitored, but no photodesorption was detected for this species. 

\section{Discussion}\label{sec:disc}

\subsection{SO photodesorption mechanism}
When photodesorption of photoproducts is measured in the laboratory, the evolution of the measured yields with the irradiated fluence can be related to the underlying dominant photodesorption mechanism, according to \citet{martin15,martin16,martin18}. 
In particular, an increasing photodesorption yield with fluence is indicative of photodesorption taking place through an energy transfer mechanism. In that process VUV photons are absorbed by a molecule in the subsurface of the ice, and then the energy is transferred to a photoproduct previously formed on the surface of the ice, that subsequently desorbs to the gas phase. Accumulation of photoproduct molecules on the ice surface as irradiation of the ice sample proceeds leads to a larger number of these molecules available for photodesorption at higher irradiated fluences, and thus to an increase in the photodesorption yield. 
A constant photodesorption yield with fluence, on the other hand, suggests that photodesorption is proceeding through the so-called photochemical desorption (or photochemidesorption) mechanism. This mechanism is characterized by an immediate desorption of the photoproduct molecule right after its formation on the ice surface,  thanks to the excess energy provided by the photon to the different fragments after photodissociation of the absorbing molecule and/or the exothermicity of the formation reaction. Since photochemidesorption takes place immediately after the formation of photoproducts, accumulation of photoproducts on the surface of the ice does not affect the corresponding photodesorption yields, leading to a constant value with the irradiated fluence. 
In Experiments 1$-$14 the only photoproduct observed to photodesorb was SO. Figure \ref{fig:so_QMS_corr} shows that the photodesorption yield did not significantly change with the irradiation time. Therefore, the observed evolution of the SO photodesorption yield with the irradiated fluence was consistent with that expected from a photochemidesorption mechanism. 

We note that the estimated SO photodesorption yield in this work (4$-$8 $\times$ 10$^{-5}$ mol. ph.$^{-1}$) corresponds to SO molecules forming and subsequently photodesorbing from a SO$_2$-rich ice. This value could be different in irradiated SO-rich ices. 
For example, the CO photodesorption yield in irradiated CO$_2$ ice samples \citep[up to 1.3 $\times$ 10$^{-2}$ mol. ph.$^{-1}$,][]{martin15} is at least a factor of $\sim$5 lower than the measured yield in irradiated CO ices \citep[5.4 $\times$ 10$^{-2}$ mol. ph.$^{-1}$,][]{munozcaro10}. 
Unfortunately, SO cannot be purchased for laboratory experiments because it is not a stable molecule on Earth. Therefore, it is not possible to grow a pure SO ice sample in the laboratory to measure the photodesorption yield under those conditions.

\subsection{Contribution of ice photodesorption to the SO$_2$ and SO gas-phase abundances in the Horsehead PDR}\label{sec:disc_HH}

The estimated SO$_2$ photodesorption yield is comparable to that measured for CO$_2$ molecules upon VUV irradiation of CO$_2$ ice samples \citep[$\sim$1 $\times$ 10$^{-4}$ mol. ph.$^{-1}$,][]{martin15}, whereas it is between one and two orders of magnitude lower than those measured for NH$_3$ and CO upon VUV irradiation of NH$_3$ and CO ice samples \citep[respectively]{martin18,munozcaro10}. 
In all cases, broadband VUV irradiation was performed with the same MDHL. 
This illustrates the wide range of photodesorption yields that can be measured for different species, and highlights the fact that photodesorption is a molecule-specific process.  


\begin{figure*}
    \centering
    \includegraphics[width=\linewidth]{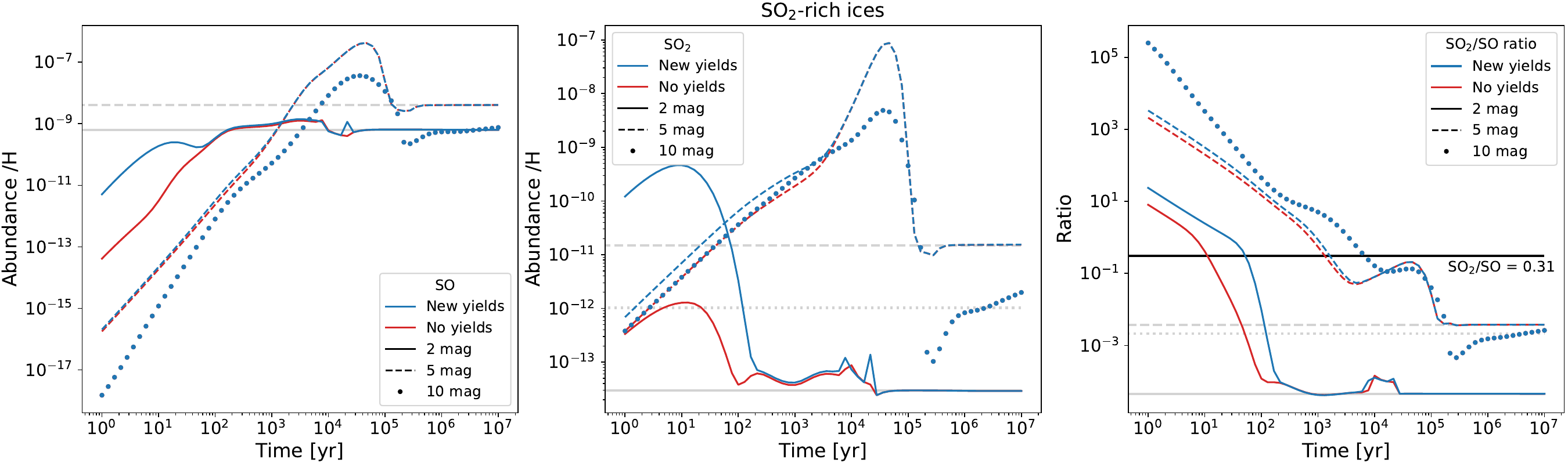}
    \caption{Model predictions of SO (left panel), SO$_{2}$ (middle panel), and the SO$_{2}$/SO ratio (right panel) in the SO$_{2}$-rich ice scenario. The predictions that include our photodesorption yields are plotted in blue, while for the predictions in red the photodesorption yields were set to zero. Predictions at different extinctions are depicted with varying line styles: 2 mag (solid line), 5 mag (dashed line), and 10 mag (dotted line). The steady state values are represented with horizontal gray lines. The SO$_{2}$/SO ratio observed toward the Horsehead PDR is shown as an horizontal black line in the right panel.}
    \label{fig:SO2richIce}
\end{figure*}

In this Section, we explore the impact of SO$_2$ and SO photodesorption in an astrophysical environment.  
Photon-dominated regions (PDRs) are regions of the ISM where far-ultraviolet (FUV) photons are the main drivers of heating and chemical processes, thus representing a suitable environment to test the impact of the derived photodesorption yields. 
Among the PDRs found in our Galaxy, the Horsehead is one of the most well-known examples of low-UV PDRs \citep{Abergel2003}. Due to its proximity \citep[$\sim 400$ pc,][]{AnthonyTwarog1982} and simple geometry, it has been the target of numerous chemical studies, unveiling a rich molecular chemistry in sulfur-bearing compounds \citep[see, e.g.,][]{Goicoechea2006, fuente17, RiviereMarichalar2019}, small hydrocarbons \citep{Guzman2015}, and complex organic molecules \citep[see, for instance,][]{Gratier2013, Guzman2013, Guzman2014, LeGal2017}. The moderate FUV illumination and the detection of S$_{2}$H \citep{fuente17} suggest that ice mantles are indeed being photoprocessed in this PDR.  

In the following, we present a static isobaric toy model of the Horsehead PDR (Appendix \ref{sec:appHH}) coupled with the time dependent, gas-grain chemical model \texttt{Nautilus} \citep{Ruaud2016} to evaluate the contribution of the SO$_2$ and SO photodesorption yields to their predicted gas-phase abundances, as well as to the SO$_{2}$/SO ratio.
\texttt{Nautilus} computes the chemical composition over time given an initial set of physical and chemical conditions. 
This code  takes into account the chemical processes and interactions in the gas-phase, the surface of icy grains, and the bulk of ice mantles in a particular interstellar region. We used the most up-to-date version \texttt{Nautilus 2.0.0} with its corresponding updated chemical network, described in \citet{Wakelam2024}.
In order to more easily analyze the effect of SO$_2$ and SO photodesorption, we considered as a first approximation a SO$_2$-rich ice scenario, where the total budget of sulfur $(1.5\times 10^{-5})$ was initially in the form of SO$_{2}$ ice. 
In addition, the carbon-to-oxygen ratio was set to C/O $=0.71$, all carbon was initially locked in CO ice, and the remaining oxygen abundance was present in the gas phase.  
The resulting evolution of the SO$_2$ and SO gas-phase abundances, as well as the SO$_{2}$/SO ratio, at different extinctions are shown in Fig. \ref{fig:SO2richIce}. 
To highlight the contribution of photodesorption, Fig. \ref{fig:SO2richIce} also presents the resulting evolution when the SO$_{2}$ and SO photodesorption yields are set to zero.

The left panel of Fig. \ref{fig:SO2richIce} shows that, regardless of the assumed photodesorption yield, the SO gas-phase abundance steadily grows with time until reaching its steady state value. 
This growth is nevertheless dependent on the extinction. 
Higher exposure to UV photons (i.e., lower extinction) leads to a faster chemistry, achieving the steady-state value in $\sim100$ years. In contrast, shielded areas of the PDR require at least $\sim10^{5}$ years to reach equilibrium. 
Our estimate of the photodesorption yield had a greater influence at the lowest extinction, leading to an even faster increase of the SO gas-phase abundance during the first $\sim100$ years when the SO photodesorption is taken into account. 
In that case, the predicted growth of the SO gas-phase abundance is mainly driven by the photodissociation of SO$_{2}$ into SO + O in the gas phase combine with the photodesorption of SO from the ice mantles, while its destruction is driven by its photodissociation into S and O. 
This suggests that photodesorption of SO might be one of the main chemical processes releasing this molecule into the gas phase at extinctions $< 5$ mag. 

Similarly, the effect of the SO$_2$ photodesorption yield 
on the predicted gas-phase abundance of SO$_{2}$ 
was higher at low extinctions during the first $\sim100$ years (middle panel of Fig. \ref{fig:SO2richIce}). 
According to our chemical network, during the first 10 years the SO$_{2}$ gas-phase abundance grows due to the photodesorption of SO$_{2}$ ice, that outpaces its photodissociation. 
After reaching its peak in $\sim 10$ years, the electronic recombination reaction HSO${_{2}}^{+}$ + e$^{-}$ $\rightarrow$ H + SO$_{2}$ overcomes the photodesorption of SO$_{2}$ as the main contributor.  
It is, however, less efficient than the photodesorption of SO$_{2}$ in the first 10 years. This less efficient pathway combined with the photodissociation of SO$_{2}$ that still takes place, leads to the decline in the SO$_{2}$ gas-phase abundance until reaching the steady state value. 

As expected, the predicted SO$_{2}$/SO ratio was also affected by the derived photodesorption yields, especially at lower extinctions (right panel of Fig. \ref{fig:SO2richIce}). 
This ratio was estimated from single-dish observations toward the Horsehead at two different positions, presenting a value of 
SO$_{2}$/SO $=0.31$ at the PDR position, and SO$_{2}$/SO $=0.14$ in the shielded portion of the Horsehead where $T_{\rm k}\sim 10-20$ K \citep{RiviereMarichalar2019}. 
The physical properties of the PDR position are comparable to the physical conditions of our model at an extinction of $\sim$2 mag. 
The observed ratio (black line in the right panel of Fig. \ref{fig:SO2richIce}) is only reproduced by the model at 2 mag 
for times around $\sim$100 years, whereas it 
is much higher than the steady state value. 
This suggests possible dynamical effects not taken into account by our static toy model. 
In fact, \cite{HernandezVera2023} discussed that, even though the static isobaric approximation may be adequate for the edge of the molecular filaments found with ALMA observations toward the Horsehead, the PDR interface seems to be compressed, adding a dynamical component not taken into account in our model.

Figure \ref{fig:H2SrichIce} in Appendix \ref{sec:appHH} presents the same modeling results but considering a H$_{2}$S-rich ice scenario where the total sulfur budget is initially in H$_{2}$S ice \citep[note that H$_2$S is the most abundand S-bearing molecule detected in comet 67P/Churyumov-Gerasimenko,][]{calmonte16}. 
In the H$_{2}$S-rich ice case, the derived photodesorption yields had a limited impact on the SO$_{2}$ and SO gas-phase abundances, even at low extinctions, due to the lack of SO$_{2}$ ice in this scenario.
We also note that the observed SO$_{2}$/SO ratio at the PDR position was not well reproduced by the model in this scenario. 
In any case, the contribution of photodesorption to the gas-phase abundances across different interstellar environments (including cold, dense clouds), and for different ice compositions were beyond the scope of this paper, and will be addressed in a follow-up publication.

\subsection{Detectability of SO$_3$ in interstellar ice IR spectra}

\begin{figure}
    \centering
    \includegraphics[width=1\linewidth]{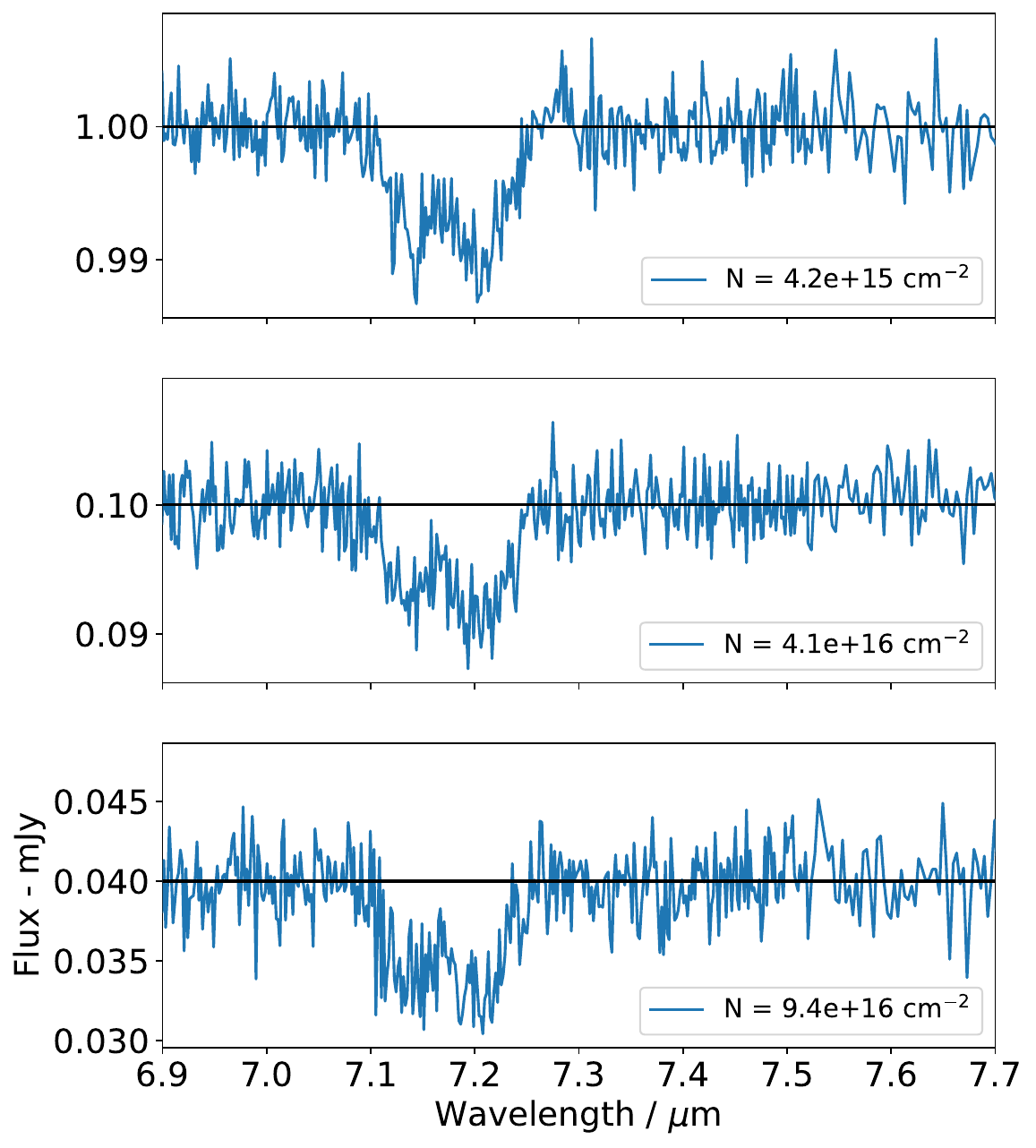}
    \caption{Detection threshold determination of SO$_3$ for 3 different flat continua (from top to bottom: 1, 0.1 and 0.04 mJy) with a fiducial noise extracted from \citet{mcclure23}. 
    The synthetic ice spectra in blue represent the absorption of the SO$_3$ vibrational mode for the column densities needed to obtain a 5$\sigma$ detection with respect to the flat continuum (in black) and the added noise.}
    \label{fig_synth}
\end{figure}

The $\sim$1395 cm$^{-1}$ ($\sim$7.17 $\mu$m) SO$_3$ IR band falls within the observing range of the MIRI instrument onboard the James Webb Space Telescope (JWST). 
Since the parent molecule SO$_2$ has been tentatively detected in interstellar ices \citep{boogert97,mcclure23}, SO$_3$ could also be detected thanks to the high sensitivity and spectral resolution of the MIRI instrument. 
In order to evaluate the detectability of this molecule in interstellar ice IR spectra, we incorporated the acquired information of the SO$_3$ IR feature (including the band strength calculated as part of this work, Table \ref{tab:ir}) into the Synthetic Ice Spectra Generator (SynthIceSpec) code.  
SynthIceSpec have been previously used in \citet{Taillard_2025a}, and it is presented in detail in \citet{Taillard_2025b}. 
The code produces synthetic absorption spectra of interstellar ices toward a particular source taking into account the JWST instrument resolution and observing modes. 
In this code, each vibrational mode is represented by a Gaussian or a sum of Gaussians, and the column density of every species is input by the user.

The detectability of SO$_3$ was evaluated using a similar method as in \citet{Taillard_2025a}. 
As a first approximation, we studied a simple case in which we did not consider the presence of any other ice molecules. 
Instead, we simply evaluated the 
detectability of the SO$_3$ IR feature using MIRI/MRS parameters toward 
three different background sources presenting a flat continuum of 1, 0.1 and 0.04 mJy (representing different levels of extinction encountered in dense clouds).
In summary, we added a fiducial noise \citep[extracted from][]{mcclure23} to the flat continua, and then iteratively increased the SO$_3$ column density to produce different synthetic SO$_3$ absorption spectra.
Finally, 5$\sigma$ detection thresholds were estimated by calculating the signal-to-noise (SNR) ratio of the SO$_3$ feature \citep[see][for more information]{Taillard_2025a}. 
The resulting detection thresholds are presented in Figure.~\ref{fig_synth}. 
The minimum column density required for a 5$\sigma$ detection was 4.2 $\times$ 10$^{15}$, 4.1 $\times$ 10$^{16}$ and 9.4 $\times$ 10$^{16}$ cm$^{-2}$ for a 1, 0.1 and 0.04 mJy flux, respectively.

We note that these estimated detection thresholds are lower than the ones derived for SO$_2$ in \citet{Taillard_2025a}. However, they do not take into account potential contributions from additional molecules, as mentioned above. 
These contributions could be expected, since this spectral region corresponds to the fingerprint region of several species. 
For example, multiple COM vibrational modes could be contributing to the IR absorption in this region, such as the HCOOH bending mode at 7.2 $\mu$m.
Most importantly, the C-H bending mode of CH$_3$OH (which has a high band strength and band width), could partly overlap with the SO$_3$ IR feature. 
In addition, uncertainties in the continuum subtraction could also have an effect on the identification and quantification of potentially weak IR bands. Therefore,
identification of this S-bearing molecule in interstellar ice spectra should thus be done with careful fitting of a local continuum, combined with a careful removal of the CH$_3$OH contribution.
On this regard, an accurate determination of the CH$_3$OH ice column density using alternative IR bands would be fundamental to calculate its contribution to the 7.2 $\mu$m band.
%
As a result, the thresholds indicated above should be carefully considered as possible lower-limits, depending on the type of source observed (and thus, its continuum), the ice composition, and other possible observational effects.

\begin{figure}
    \centering
    \includegraphics[width=1\linewidth]{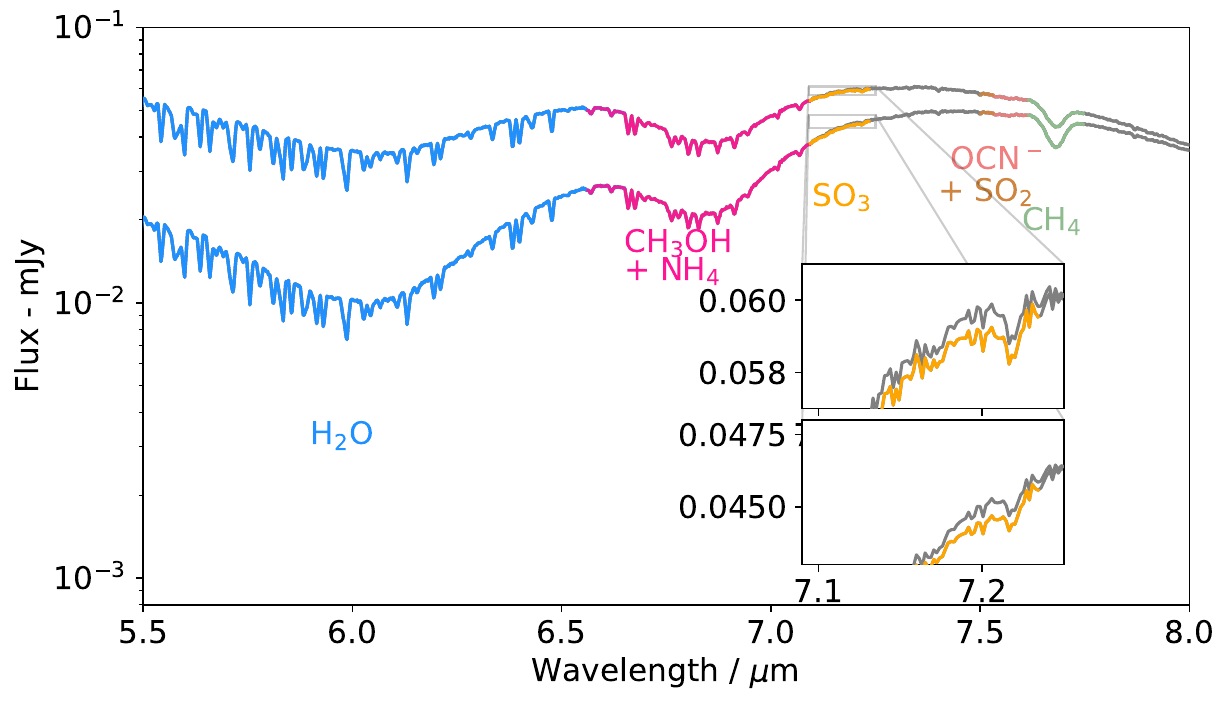}
    \caption{Synthetic ice absorption spectra toward the two background stars NIR38 and J110621 in Chameleon I. 
    The background stars are represented by a K7V stellar spectrum (in grey), according to \citet{Dartois_2024}. 
    Absorption features corresponding to different ice species are highlighted in color. Ice column densities were extracted from \citet{mcclure23}. 
    Insets show the spectra without the SO$_3$ feature (in grey) , and with the SO$_3$ column density required for a 5$\sigma$ detection derived in this work (in yellow). 
    }
    \label{fig_synth_mcclure}
\end{figure}

In order to evaluate the detectability of SO$_3$ in a more realistic scenario, we included the column densities of all ice species listed in Table 2 of \citet{mcclure23} for the two different reported observations (background stars NIR38 and J110621) as inputs in SynthIceSpec. 
We also included a photostellar spectrum and extinction in both cases, corresponding to a K7V stellar type (T$_{eff}$ = 3900 K). 
This spectrum was considered as the best fit to the observations in \citet{Dartois_2024}. 
We then carried out the same iterative process as explained above. 
The resulting synthetic absorption spectra corresponding to both background stars are shown in Fig.~\ref{fig_synth_mcclure}. 
In a more realistic scenario, 5$\sigma$ detection of the SO$_3$ feature would be possible for column densities of about 4.5 $\times$ 10$^{16}$ and 5.5 $\times$ 10$^{16}$ cm$^{-2}$ for NIR38 and J110621, respectively. 
These values are slightly higher than the column densities proposed for SO$_2$ in \citet{mcclure23} (3.4 $\times$ 10$^{16}$ and 4.7 $\times$ 10$^{16}$ cm$^{-2}$, respectively). 
%
Even though the detection threshold is too high in this particular case, SO$_3$ should still be considered as a potential contributor to the $\sim$7.2 $\mu$m absorption in other scenarios, as its band strength is strong. 

\section{Conclusions}\label{sec:conc}
We have experimentally measured the photodesorption yields of SO$_2$ and, tentatively, SO molecules upon VUV photon irradiation of SO$_2$ ice samples at 14$-$80 K:  
\begin{itemize}
    \item The estimated photodesorption yields at 14 K were $\sim$2.3 $\times$ 10$^{-4}$ mol. ph.$^{-1}$ and $\sim$6 $\times$ 10$^{-5}$ mol. ph.$^{-1}$ for SO$_2$ and SO, respectively.
    \item The SO$_2$ photodesorption yield increased with temperature between 22 K and 70 K, up to a value of 3.8 $\times$ 10$^{-4}$ mol. ph.$^{-1}$ at 70 K. At the same time, we observed slightly lower yields for crystalline ices compared to amorphous ices irradiated at similar temperatures. 
    \item No clear trend with temperature was observed for the SO photodesorption yield. The evolution of the photodesorption yield with the irradiated fluence suggested that SO molecules photodesorbed immediately after their formation through a photochemical desorption mechanism.  
    \item According to astrochemical models, the impact of the estimated SO$_2$ and SO photodesorption yields in their gas-phase abundances in PDRs would be larger at extinctions $<$ 5 mag. The contribution of photodesorption across different environments will be addressed in a follow-up paper.   
    \item The main detected product in irradiated SO$_2$ ices was SO$_3$. We theoretically calculated the band strength of the $\sim$1395 cm$^{-1}$ SO$_3$ IR feature and obtained a value of 1.1 $\times$10$^{-16}$ cm molecule$^{-1}$. 
    With this band strength, ice column densities of at least 4.2 $\times$ 10$^{15}$, 4.1 $\times$ 10$^{16}$ and 9.4 $\times$ 10$^{16}$ cm$^{-2}$ would be needed for a 5$\sigma$ detection of this species in interstellar ice IR spectra observed toward a background source with a flat continuum of 1, 0.1 and 0.04 mJy, respectively. 
\end{itemize}

The derived photodesorption yields represent a key constraint for astrochemical models aiming to reproduce the gas-phase abundances of sulfur-bearing species, allowing a more accurate interpretation of observations of (especially) cold regions of the ISM. 
In addition, the estimated SO$_3$ IR band strength could contribute to the potential identification and quantification of this species in IR spectra obtained with JWST, improving our understanding of the sulfur reservoirs in star-forming regions.

\section*{Acknowledgements}
The project leading to these results has received funding from “la Caixa” Foundation, under agreement LCF/BQ/PI22/11910030. 
BE acknowledges the support by grant PTA2020-018247-I from the Spanish Ministry of Science and Innovation/State Agency of Research MCIN/AEI. 
DN, AT, and AF have received funding from the European Research Council (ERC) under the European Union’s Horizon Europe research and innovation programme ERC-AdG-2022 (GA No. 101096293). 
GMMC and HC received funding from project PID2020-118974GB-C21 by the Spanish Ministry of Science and Innovation.
\section*{Data Availability}
The experimental data underlying this article are available in the \texttt{zenodo} repository at https://zenodo.org, and can be accessed with DOI: 10.5281/zenodo.15690762. 

\appendix




\begin{thebibliography}{99}
\bibitem[\protect\citeauthoryear{Abergel et al.}{2003}]{Abergel2003} Abergel A., et al., 2003, A\&A, 410, 577
\bibitem[\protect\citeauthoryear{Anthony-Twarog}{1982}]{AnthonyTwarog1982} Anthony-Twarog B.J., 1982, AJ, 87, 1213
\bibitem[\protect\citeauthoryear{Artur de la Villarmois et al.}{2018}]{villarmois18} Artur de la Villarmois E., 
et al., 2018, A\&A, 614, A26
\bibitem[\protect\citeauthoryear{Artur de la Villarmois et al.}{2019}]{villarmois19} Artur de la Villarmois E., J{\o}rgensen J.K., Kristensen L.E., Bergin E.A., Harsono D., Sakai N., van Dishoeck E.F., Yamamoto S., 2019, A\&A, 626, A71
\bibitem[\protect\citeauthoryear{Bertin et al.}{2012}]{bertin12} Bertin M., et al.,2012, Phys. Chem. Chem. Phys., 14, 9929 
\bibitem[\protect\citeauthoryear{Boissier et al.}{2007}]{boissier07} Boissier J., 
et al., 2007, A\&A, 475, 1131
\bibitem[\protect\citeauthoryear{Boogert et al.}{1997}]{boogert97} Boogert A.C.A., Schutte W.A., Helmich F.P., Tielens A.G.G.M., Wooden D.H., 1997, A\&A, 317, 929
\bibitem[\protect\citeauthoryear{Booth et al.}{2023}]{booth23} Booth A., Ilee J.D., Walsh C., Kama M., Keyte L., van Dishoeck E.F., Hideko N., A\&A, 669, A53
\bibitem[\protect\citeauthoryear{Booth et al.}{2021}]{booth21} Booth A., van der Marel N., Leemker M., van Dishoeck E.F., Ohashi S., 2021, A\&A, 651, L6
\bibitem[\protect\citeauthoryear{Booth et al.}{2018}]{booth18} Booth A., Walsh C., Kama M., Loomis R.A., Maud L.T., Juh\'asz A., 2018, A\&A, 611, A16
\bibitem[\protect\citeauthoryear{Bron et al.}{2018}]{Bron2018} Bron E., Ag\'undez M., Goicoechea J.R., Cernicharo J., 2018, arXix e-prints, p. arXiv:1801.01547
\bibitem[\protect\citeauthoryear{Burke et al.}{2005}]{burke05} Burke D.J., Vondrak T., Meech S.R., 2005, Surf. Sci., 585, 123
\bibitem[\protect\citeauthoryear{Calmonte et al.}{2016}]{calmonte16} Calmonte U. et al., 2016, MNRAS, 462, 1, S253
\bibitem[\protect\citeauthoryear{Cecchi-Pestellini \& Aiello}{1992}]{cecchi92} Cecchi-Pestellini C., \& Aiello S. 1992, MNRAS 258, 125
\bibitem[\protect\citeauthoryear{Chaabouni et al.}{2000}]{chaabouni00} Chaabouni H., Schriver-Mazzuoli L., Schriver A., 2000, J. Phys. Chem. A, 104, 3498
\bibitem[\protect\citeauthoryear{Charnley}{1997}]{charnley97} Charnley S.B., 1997, ApJ, 481, 396
\bibitem[\protect\citeauthoryear{Chen et al.}{2014}]{chen14} Chen Y.-J., Chuang K.-J., Mu\~noz Caro G.M., Nuevo, M., Chu C.-C., Yih T.S., Ip W.-H., Wu C.-Y. R., 2014, ApJ, 781, 15
\bibitem[\protect\citeauthoryear{Clark et al.}{2005}]{CASTEP} Clark S.J., Segall M.D., Pickard C.J., Hasnip P.J., Probert M.J., Refson K., Payne M.C., 2005, Z. Kristall., 220, 567--570
\bibitem[\protect\citeauthoryear{Cruz-D\'iaz et al.}{2016}]{gustavo16} Cruz-D\'iaz G.A., Mart\'in-Dom\'enech R., Mu\~noz Caro G.M., Chen Y.-J., 2016, A\&A, 592, A68
\bibitem[\protect\citeauthoryear{Cruz-D\'iaz et al.}{2014a}]{gustavo14a} Cruz-D\'iaz G.A., Mu\~noz Caro G.M., Chen Y.-J., Yih T.-S., 2014a, A\&A, 562, A119
\bibitem[\protect\citeauthoryear{Cruz-D\'iaz et al.}{2014b}]{gustavo14b} Cruz-D\'iaz G.A., Mu\~noz Caro G.M., Chen Y.-J., 2014c, MNRAS, 439, 2370
\bibitem[\protect\citeauthoryear{Cruz-D\'iaz et al.}{2014c}]{gustavo14c} Cruz-D\'iaz G.A., Mu\~noz Caro G.M., Chen Y.-J., Yih T.-S., 2014c, A\&A, 562, A120
\bibitem[\protect\citeauthoryear{Dalton et al.}{2010}]{dalton10} Dalton J.B., Cruikshank D.P., Stephan K., McCord T.B., Coustenis A., Carlson R.W., Coradini A., 2010, Chemical Composition of Icy Satellite Surfaces, 153
\bibitem[\protect\citeauthoryear{Dartois et al.}{2024}]{Dartois_2024} Dartois E., et al., 2024, NatAs, 8, 359
\bibitem[\protect\citeauthoryear{de Laeter et al.}{2003}]{laeter03} de Laeter J.R., B\"ohlke, J.R., De Bievre, P., et al.\ 2003, Pure Appl. Chem., 75, 685
\bibitem[\protect\citeauthoryear{de Souza Bonfim et al.}{2017}]{deSouza17} de Souza Bonfim V., de Castilho R.B., Baptista L., Pilling S., 2017, PCCP, 19, 26906
\bibitem[\protect\citeauthoryear{del Burgo et al.}{2024}]{carlos24} del Burgo Olivares C., Carrascosa H., Escribano B., Mu\~noz Caro G.M., Mart\'in-Dom\'enech R., 2024, MNRAS, 527-3, 8829
\bibitem[\protect\citeauthoryear{Dols et al.}{2024}]{dols24} Dols V., Paterson W.R., Bagenal F., 2024, J. Geophys. Res: Sp. Phys., 129, e2023JA031763 \bibitem[\protect\citeauthoryear{Dutrey et al.}{2011}]{dutrey11} Dutrey A., 
et al., 2011, A\&A, 535, A104 
\bibitem[\protect\citeauthoryear{el Akel et al.}{2022}]{elakel22} el Akel M., Kristensen L.E., Le Gal R., van der Walt S.J., Pitts R.L., Dulieu F., A\&A, 659, A100
\bibitem[\protect\citeauthoryear{Escribano et al.}{}]{bruno25} Escribano B. et al., \textit{in prep.}
\bibitem[\protect\citeauthoryear{Facchini et al.}{2021}]{facchini21} Facchini S., Teague R., Bae J., Benisty M., Keppler M., Isella A., 2021, AJ, 162-3, id.99
\bibitem[\protect\citeauthoryear{Ferrante et al.}{2008}]{ferrante08} Ferrante R.F., Moore M.H., Spiliotis M.M., Hudson R.L., 2008, ApJ, 684, 1210
\bibitem[\protect\citeauthoryear{Fillion et al.}{2014}]{fillion14} Fillion J.-H., et al., 2014, Farady Discuss., 168, 533
\bibitem[\protect\citeauthoryear{Fuente et al.}{2010}]{fuente10} Fuente A., Cernicharo J., Ag\'undez M., Bern\'e O., Goicoechea J.R., Alonso-Albi T., Marcelino N., 2010, A\&A, 524, A19
\bibitem[\protect\citeauthoryear{Fuente et al.}{2017}]{fuente17} Fuente A., et al., 2017, ApJL, 851-2, L49
\bibitem[\protect\citeauthoryear{Garozzo et al.}{2008}]{garozzo08} Garozzo M., Fulvio D., Gomis O., Palumbo M.E., Strazzulla G., 2008, Plan. and Sp. Sci., 56, 1300
\bibitem[\protect\citeauthoryear{Gerakines et al.}{1995}]{gerakines95} Gerakines P.A., Schutte W.A., Greenberg J.M., van Dishoeck E.F., 1995, A\&A, 296, 810
\bibitem[\protect\citeauthoryear{Goicoechea \& Shalyapin}{2016}]{Goicoechea2016} Goicoechea L.J., Shalyapin V.N., 2016, A\&A, 596, A77
\bibitem[\protect\citeauthoryear{Goicoechea et al.}{2006}]{Goicoechea2006} Goicoechea J.R., Pety J., Gerin M., Teyssier D., Roueff E., Hily-Blant P., Baek S., 2006, A\&A, 456, 565
\bibitem[\protect\citeauthoryear{Goicoechea et al.}{2009}]{Goicoechea2009} Goicoechea J.R., Pety J., Gerin M., Hily-Blant P., Le Bourlout J., 2009, A\&A, 498, 771
\bibitem[\protect\citeauthoryear{Gratier et al.}{2013}]{Gratier2013} Gratier P., Pety J., Guzm\'an V., Gerin M., Goicoechea J.R., Roueff E., Faure A., 2013, A\&A, 557, A101
\bibitem[\protect\citeauthoryear{Guzm\'an et al.}{2013}]{Guzman2013} Guzm\'an V.V., et al., 2013, A\&A, 560, A73
\bibitem[\protect\citeauthoryear{Guzm\'an et al.}{2014}]{Guzman2014} Guzm\'an V.V., Pety J., Gratier P., Goicoechea J.R., Gerin M., Roueff E., Le Petit F., Le Bourlot J., 2014, Faraday Discussions, 168, 103
\bibitem[\protect\citeauthoryear{Guzm\'an et al.}{2015}]{Guzman2015} Guzm\'an V.V., Pety J., Goicoechea J.R., Gerin M., Roueff E., Gratier P., \"Oberg K.I., 2015, ApJ, 800, L33
\bibitem[\protect\citeauthoryear{Gonz\'alez-D\'iaz et al.}{2022}]{cristobal22} Gonz\'alez-D\'iaz C., Carrascosa H., Mu\~noz Caro G.M., Satorre M.\'A., Chen Y.-J., 2022, MNRAS, 517, 5744
\bibitem[\protect\citeauthoryear{Gredel et al.}{1989}]{gredel89} Gredel R., Lepp S., Dalgarno A., Herbst E., 1989, ApJ, 347, 289
\bibitem[\protect\citeauthoryear{Grimme}{2006}]{SEDC-G06} Grimme S., 2006, J. Comput. Chem., 27, 1787
\bibitem[\protect\citeauthoryear{Guilloteau et al.}{2016}]{guilloteau16} Guilloteau S., 
et al., 2016, A\&A, 592, A124
\bibitem[\protect\citeauthoryear{Hehenberg \& Kohn}{1964}]{hohenberg64} Hohenberg P., Kohn W., 1964, Phys. Rev., 136, B864
\bibitem[\protect\citeauthoryear{Hern\'andez-Vera et al.}{2023}]{HernandezVera2023} Hern\'andez-Vera C., et al., 2023, A\&A, 677, A152
\bibitem[\protect\citeauthoryear{Herron \& Schiff}{1956}]{herron56} Herron J.T. \& Schiff H.I., 1956, J. Chem. Phys., 24, 1266
\bibitem[\protect\citeauthoryear{Holtom et al.}{2006}]{holtom06} Holtom P.D., Dawes A., Mukerji R.J., Davis M.P., Webb S.M., Hoffman S.V., Mason N.J., 2006, PCCP, 8, 714
\bibitem[\protect\citeauthoryear{Joblin et al.}{2018}]{Joblin2018} Joblin C., et al., 2018, A\&A 615, A129
\bibitem[\protect\citeauthoryear{Kaiser et al.}{1995}]{kaiser95} Kaiser R.I., Jansen P., Petersen K., Roessler K., 1995, Rev. Sci. Instrum., 66, 5226
\bibitem[\protect\citeauthoryear{Kaufman et al.}{1999}]{Kaufman1999} Kaufman M.J., Wolfire M.G., Hollenbach D.J., Luhman M.L., 1999, ApJ, 527, 795
\bibitem[\protect\citeauthoryear{Kohn \& Sham}{1965}]{kohn65} Kohn W., Sham L.J., 1965, Phys. Rev., 140, A1133
\bibitem[\protect\citeauthoryear{Laas \& Caselli}{2019}]{laas19} Laas J., Caselli P., 2019, A\&A, 624, A108
\bibitem[\protect\citeauthoryear{Le Gal et al.}{2021}]{legal21} Le Gal R., 
et al., 2021, ApJSS, 257-1, id.12
\bibitem[\protect\citeauthoryear{Le Gal et al.}{20217}]{LeGal2017} Le Gal R., Herbst E., Dufuor G., Gratier P., Ruaud M., Vidal T.H.G., Wakelam V., 2017, A\&A, 605, A88
\bibitem[\protect\citeauthoryear{Le Petit et al.}{2006}]{LePetit2006} Le Petit F., Nehm\'e C., Le Bourlot J., Roueff E., 2006, ApJS, 164, 506
\bibitem[\protect\citeauthoryear{Le Roy et al.}{2015}]{leroy15} Le Roy L., 
et al., 2015, A\&A, 583, A1
\bibitem[\protect\citeauthoryear{Maillard et al.}{2021}]{Maillard2021} Maillard V., Bron E., Le Petit F., 2021, A\&A 656, A65
\bibitem[\protect\citeauthoryear{Maity \& Kaiser}{2013}]{maity13} Maity S. \& Kaiser R.I., 2013, ApJ, 773, 184
\bibitem[\protect\citeauthoryear{Mart\'in-Dom\'enech et al.}{2018}]{martin18} Mart\'in-Dom\'enech R., Cruz-D\'iaz G.A., Mu\~noz Caro G.M., 2018, MNRAS, 473, 2575
\bibitem[\protect\citeauthoryear{Mart\'in-Dom\'enech et al.}{2015}]{martin15}Mart\'in-Dom\'enech R., Manzano-Santamar\'ia J., Mu\~noz Caro G.M., Cruz-D\'iaz G.A., Chen Y.-J., Herrero V.J., Tanarro I., 2015, A\&A, 584, A14
\bibitem[\protect\citeauthoryear{Mart\'in-Dom\'enech et al.}{2016}]{martin16} Mart\'in-Dom\'enech R., Mu\~noz Caro G.M., Cruz-D\'iaz G.A., 2016, A\&A, 589, A107
\bibitem[\protect\citeauthoryear{McClure et al.}{2023}]{mcclure23} McClure M.K. et al., 2023, NatAs, 7, 431
\bibitem[\protect\citeauthoryear{Moore}{1984}]{moore84} Moore M.H., 1984, Icarus, 59, 114
\bibitem[\protect\citeauthoryear{Moore et al.}{2007}]{moore07} Moore M., Hudson R.L., Carlson R.W., 2007, Icarus, 189, 409
\bibitem[\protect\citeauthoryear{Mozo et al.}{2007}]{mozo07} Mozo R., Agusta M.K., Rahman M.M., A Diño W., Rodulfo E., Kasai H., 2007, Jour. of Phys.: Cond. Matter, 19, 36, 365244
\bibitem[\protect\citeauthoryear{Mifsud et al.}{2022}]{mifsud22} Mifsud D.V. et al., 2022, Front. in Chem., 10, id.1003163
\bibitem[\protect\citeauthoryear{Mu\~noz Caro et al.}{2010}]{munozcaro10} Mu\~noz Caro G.M., et al., 2010, A\&A, 522, A108
\bibitem[\protect\citeauthoryear{\"Oberg et al.}{2007}]{oberg07} \"Oberg K.I., Fuchs G.W., Awad Z., Fraser H.J., Schlemmer S., van Dishoeck E.F., Linnartz H., 2007, 662-1, L23
\bibitem[\protect\citeauthoryear{Perdew et al.}{1996}]{PBE} Perdew J.P., Burke K., Ernzerhof M., 1996, Phys. Rev. Lett., 77, 3865--3868
\bibitem[\protect\citeauthoryear{Podio et al.}{2015}]{podio15} Podio L., Codella C., Gueth F., et al., 2015, A\&A, 581, A85
\bibitem[\protect\citeauthoryear{Pollack et al.}{1980}]{pollack80} Pollack J.B., Toon O.B., Boese R., 1980, J. Geophys. Res., 85, 8223
\bibitem[\protect\citeauthoryear{Refson et al.}{2006}]{CASTEP_DFPT} Refson K., Tulip P.R., Clark S.J., 2006, Phys. Rev. B., 73, 155114
\bibitem[\protect\citeauthoryear{Rejoub et al.}{2002}]{rejoub02} Rejoub R., Lindsay B.G., Stebbings R.F. 2002, Phys. Rev. A, 65, 042713
\bibitem[\protect\citeauthoryear{Rivière-Marichalar et al.}{2019}]{RiviereMarichalar2019} Rivière-Marichalar P., et al., 2019, A\&A, 628, A16
\bibitem[\protect\citeauthoryear{Ruaud et al.}{2016}]{Ruaud2016} Ruaud M., Wakelam V., Hersant F., 2016, MNRAS, 459, 3756
\bibitem[\protect\citeauthoryear{Sakai et al.}{2016}]{sakai16} Sakai N., Oya Y., L\'opez-Sepulcre A., et al., 2016, ApJL, 820, L34
\bibitem[\protect\citeauthoryear{Sakai et al.}{2014a}]{sakai14a} Sakai N., Oya Y., Sakai T., et al., 2014a, ApJL, 791, L38
\bibitem[\protect\citeauthoryear{Sakai et al.}{2014b}]{sakai14b} Sakai N., Sakai T., Hirota T., et al., 2014b, Nature, 507, 78
\bibitem[\protect\citeauthoryear{Schriver-Mazzuoli et al.}{2003a}]{schriver03a} Schriver-Mazzuoli L., Chaabouni H., Schriver A., 2003a, J. Mol. Str., 644, 151
\bibitem[\protect\citeauthoryear{Schriver-Mazzuoli et al.}{2003b}]{schriver03} Schriver-Mazzuoli L., Schriver A., Chaabouni H., 2003b, Can. J. Phys., 81, 301
\bibitem[\protect\citeauthoryear{Semenov et al.}{2018}]{semenov18} Semenov D., Favre C., Fedele D., et al., 2018, A\&A, 617, A28
\bibitem[\protect\citeauthoryear{Shen et al.}{2004}]{shen04}Shen C. J., Greenberg J. M., Schutte W. A., van Dishoeck E. F., 2004, A\&A, 415, 203
\bibitem[\protect\citeauthoryear{Taillard et al.}{2025a}]{Taillard_2025a} Taillard A., et al., 2025a, A\&A, 694, A263
\bibitem[\protect\citeauthoryear{Taillard et al.}{2025b}]{Taillard_2025b} Taillard A., et al., 2025b, A\&A, \textit{submitted}
\bibitem[\protect\citeauthoryear{van der Tak et al.}{2003}]{vandertak03} van der Tak F.F.S., Boonman A.M.S., Braakman R., van Dishoeck E:F., 2003, A\&A, 412, 133
\bibitem[\protect\citeauthoryear{Wakelam et al.}{2024}]{Wakelam2024} Wakelam V., Gratier P., Loison J.C., Hickson K.M., Penguen J., Mechineau A., 2024, A\&A, 689, A63
\bibitem[\protect\citeauthoryear{Wesemberg et al.}{2007}]{wesemberg07} Wesemberg C., Autzen O., Hasselbrink E., 2007, Appl. Phys. A, 88, 559
\bibitem[\protect\citeauthoryear{Yarnall \& Hudson}{2022}]{yarnall22} Yarnall Y.Y., Hudson R.L., 2022, ApJL, 931, L4
\bibitem[\protect\citeauthoryear{Yung \& Demore}{1982}]{yung82} Yung Y.L., Demore W.B., 1982, Icarus, 51, 199

\bibitem[\protect\citeauthoryear{}{}]{}

\end{thebibliography}




\appendix

\section{Calibration of the QMS}\label{sec:appA}

\subsection{Calibration of $k_{CO}$}
\begin{figure*}
    \centering
    \includegraphics[width=0.4\linewidth]{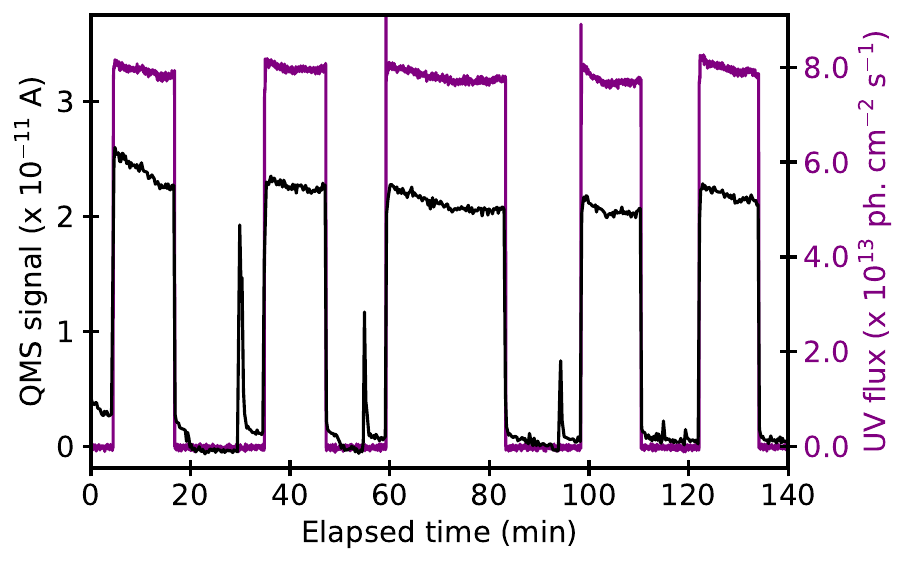}
    \includegraphics[width=0.3\linewidth]{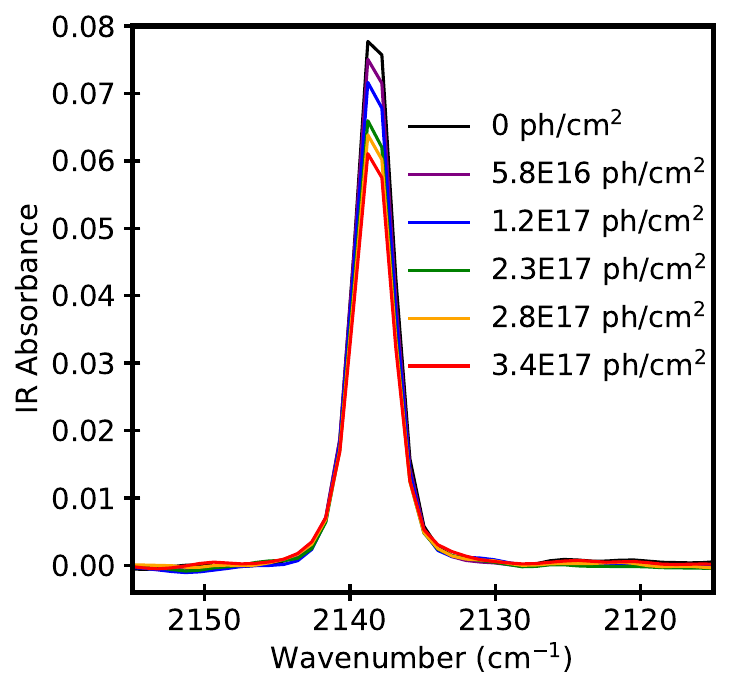}
\caption{\textit{Left panel}: Measured ion current for the $m/z$=28 signal corresponding to photodesorbing CO molecules (black line), along with the measured UV photon flux (purple line), during irradiation of a pure CO ice.
\textit{Right panel}: IR band corresponding to CO during irradiation of a pure CO ice. Different lines correspond to different irradiated fluences.
}
    \label{fig:co_QMS}
\end{figure*}


In order to estimate the photodesorbing column densities using Eq. \ref{eqn_qms}, we performed a calibration experiment to calculate the $k_{CO}$ proportionality constant \citep{martin15}. 
The calibration experiment consisted of the UV-irradiation of a pure CO ice sample made by deposition of CO (gas, Nippon 99.998\%). 
According to Eq. \ref{eqn_qms}, the $k_{CO}$ proportionality constant could be calculated in this experiment as:

\begin{equation} 
k_{CO} = \frac{I(28)}{N(CO)},  \label{eq_kco}
\end{equation}

\noindent where $I(28)$ was the integrated ion current of the $m/z$=28 signal for every irradiation interval (left panel of Fig. \ref{fig:co_QMS}), and $N(CO)$ was the corresponding photodesorbed column density. 
Since photon-induced chemistry in UV-irradiated CO ices could be considered negligible \citep{munozcaro10}, the photodesorbed CO column density in every irradiation interval could be calculated using Eq. \ref{eqn} from the difference in the integrated absorbance of the $\sim$2140 cm$^{-1}$ IR feature before and after irradiation (right panel of Fig. \ref{fig:co_QMS}). 
A 8.7 $\times$ 10$^{-18}$ cm molecule$^{-1}$ CO IR band strength was reported in \citet{cristobal22}, with a 6\% uncertainty. 
We calculated a $k_{CO}$ value for every irradiation interval in the calibration experiment using Eq. \ref{eq_kco}. The mean $k_{CO}$ value for the five irradiation intervals was 9.2 $\times$ 10$^{-11}$ A min ML$^{-1}$, with a 20\% standard deviation. 
We considered that the 6\% uncertainty in the CO IR band strength was negligible compared to the measured standard deviation, and assumed a systematic uncertainty of 20\% for $k_{CO}$.

As mentioned in Sect. \ref{sec:exp_qms}, the $k_{CO}$ proportionality constant changed from one experiment to another. 
We calculated the $k_{CO}$ values corresponding to Experiments 1$-$14 by multiplying the value derived from the calibration experiment (1.2 $\times$ 10$^{-10}$ A min ML$^{-1}$) by a correction factor. 
The correction factor was calculated as the ratio of the integrated $m/z$=64 QMS signal during deposition of the SO$_2$ ice in Experiments 1$-$14 (and normalized by the deposited SO$_2$ ice column density) with respect to the value measured the day the $k_{CO}$ calibration experiment was performed. 
Note that the integrated $m/z$=64 QMS signal measured during deposition and normalized by the resulting SO$_2$ column density was expected to be the same for all experiments. 
Therefore, differences across experiments were due to changes in the proportionality constant $k_{CO}$ that related the measured ion current with the number of detected SO$_2$ molecules. 
The corresponding $k_{CO}$ values for each experiment are listed in the seventh column of Table \ref{tab:exp}. 

\subsection{Calibration of $S(m/z)$}
\begin{figure}
    \centering
    \includegraphics[width=0.6\linewidth]{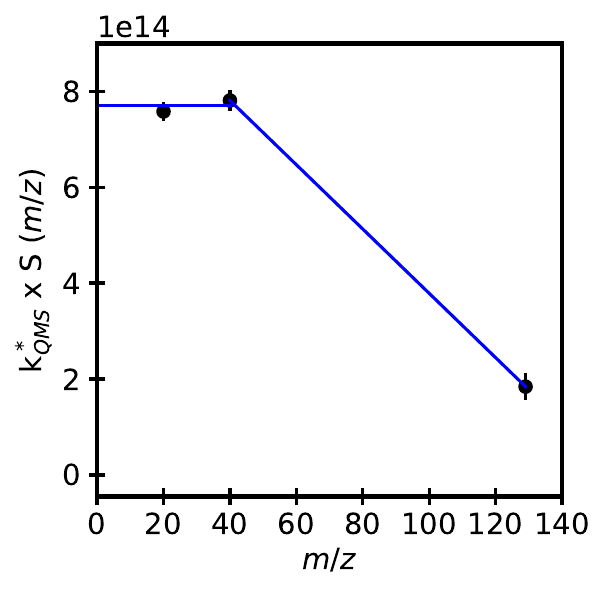}
    \caption{Dependence of the QMS sensitivity with the monitored $m/z$.}
    \label{fig:QMS_S}
\end{figure}

\begin{table}
    \centering
    \begin{tabular}{cccc}
    Factor & Ne & Ar & Xe\\
        \hline
    monitored $m/z$ & 20 & 40 & 129 \\
    $\sigma^+(X)$ (\AA$^2$)$^a$ & 0.475 & 2.520 & 4.670\\
    $I_F(z)^b$ & 0.990 & 0.921 & 0.688\\
    $Is_F(z)^c$ & 0.905 & 0.996 & 0.264\\
    \hline
    \end{tabular}
    \caption{Parameters used in Eq. \ref{eqn_qms} to calibrate the sensitivity of the QMS with $m/z$. 
    $^a$From \citet{rejoub02}.
    $^b$Measured in ISAC. 
    $^c$From \citet{laeter03}.
    }
    \label{tab:qms_param_noble}
\end{table}

The calibration of the QMS sensitivity $S(m/z)$ was performed following the protocol described in \citet{martin15}. 
To that purpose, we introduced different pressures of three noble gases into the UHV chamber at room temperature: Ne (gas, Air Liquide, 99.99\%), Ar (gas, Air Liquide, 99.999\%), and Xe (gas, Praxair, 99.999\%). 
In this case, the measured ion current ($I(m/z)$) corresponding to the atomic masses of Ne ($m/z$ = 20), Ar ($m/z$ = 40), and Xe ($m/z$ = 129) was proportional to the measured pressure in the chamber ($P(X)$):

\begin{equation}
    I(m/z) = k^*_{QMS} \cdot P(X) \cdot  \sigma^{+}(X)  \cdot I_F(z) \cdot Is_F(m) \cdot S(m/z), \label{eq_qms_noble}
\end{equation}

\noindent where $k^*_{QMS}$ was a different proportionality constant, and $I_F(z)$ and $Is_F(m)$ were an ionization fraction and isotopic fraction, respectively. 
The $Is_F(m)$ factor was included to only take into account the contribution of the isotope of mass $m$ to the total pressure in the chamber.  
Likewise, the $I_F(z)$ factor corresponded to the fraction of ionized atoms with charge $z$.  
At the same time, the $F_F$ factor was equal to 1 since atomic species cannot fragment in the QMS, and was not included in Eq. \ref{eq_qms_noble}. 
For every noble gas, a $k^*_{QMS}$ $\cdot$ $S(m/z)$ value was calculated using Eq. \ref{eq_qms_noble} and the parameters listed in Table \ref{tab:qms_param_noble}. 
Figure \ref{fig:QMS_S} shows the dependence of the QMS sensitivity with the mass of the monitored fragment. 
After applying all corrections, we did not observe a change in the sensitivity of Ne and Ar, while the sensitivity for Xe was a factor of $\sim$3 lower. 
Therefore, as a first approximation, we considered that the sensitivity was constant for $m/z$ $<$ 40 ($k^*_{QMS}$ $\cdot$ $S(m/z)$ $\approx$ 7.7 $\times$ 10 $^{14}$ A mbar$^{-1}$ \AA$^2$) and decreased linearly for $m/z$ $>$ 40 ($k^*_{QMS}$ $\cdot$ $S(m/z)$ = 1.05 $\times$ 10 $^{15}$ $-$ 6.7 $\times$ 10 $^{12}$ $\times$ $m/z$). 

\section{Estimation of the SO$_3$ band strength and its uncertainty}
\label{sec:appDFT}

\begin{figure}
    \centering
    \includegraphics[width=1.0\linewidth]{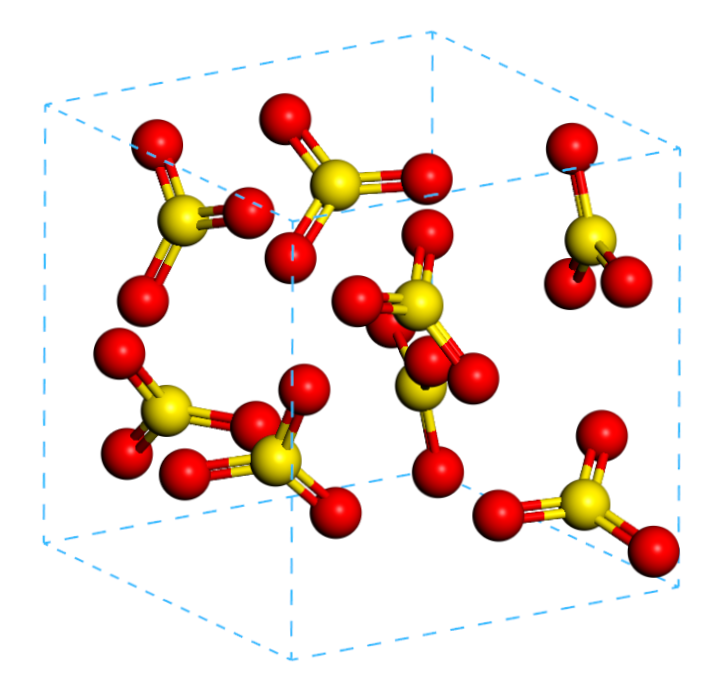} 
    \caption{Example of a simulated SO$_3$ amorphous ice with 8 molecules in a cubic box with periodic boundary conditions.}
    \label{fig:SO3_box}
\end{figure}

\begin{figure}
    \centering
    \includegraphics[width=1.0\linewidth]{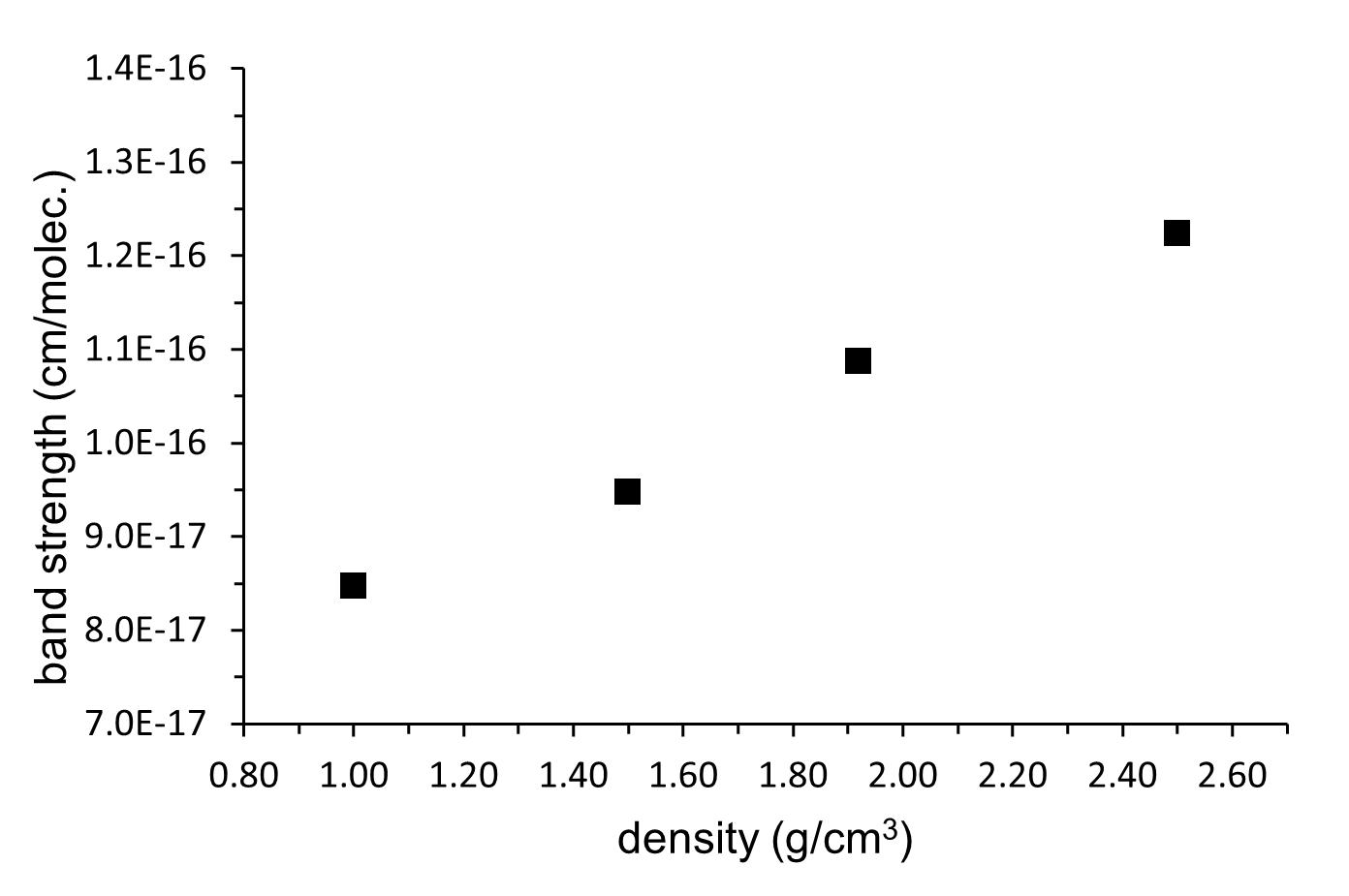} 
    \caption{Calculated IR band strengths for the anti-symmetric stretching mode of simulated SO$_3$ amorphous ice for densities between 1.0 and 2.5 g/cm$^3$.}
    \label{fig:SO3_str-dens}
\end{figure}

The method described in Sect. \ref{sec:exp_dft} to derive IR band strengths relies on previous knowledge of the ice density, which in the case of amorphous ices is not always available. 
In particular, the density of an amorphous SO$_3$ ice has not yet been experimentally determined. 
Amorphous ices often have densities close to that of their liquid phase. For example, \cite{yarnall22} reported a density of 1.395 g cm$^{-3}$ for an amorphous SO$_2$ ice, very close to the density of liquid SO$_2$ listed in the ILO database (1.4 g cm$^{-3}$).
Therefore, we assumed (as a first approximation) a density of 1.92 g cm$^{-3}$ for an amorphous SO$_3$ ice, as reported for liquid SO$_3$ in the ILO database.
The resulting amorphous SO$_3$ ice model after geometry optimization is shown in Fig. \ref{fig:SO3_box}.
%
In order to estimate the uncertainty associated to the approximation of the ice density, we calculated the band strength of the SO$_3$ anti-symmetric stretching mode for similar systems built with densities in the range 1.0 $-$ 2.5 g cm$^{-3}$. 
The results are plotted in Fig. \ref{fig:SO3_str-dens}, where a nearly linear dependence of the band strength with ice density is shown. 
If we consider a conservative error of $\pm$25\% in the assumed liquid-like density, the resulting divergence in the band strength would be approximately $\pm$15\%. 
This is the estimated uncertainty for the band strength value listed in Table \ref{tab:ir}.

\section{Additional QMS plots}
\begin{figure*}
    \centering
    \includegraphics[width=0.6\linewidth]{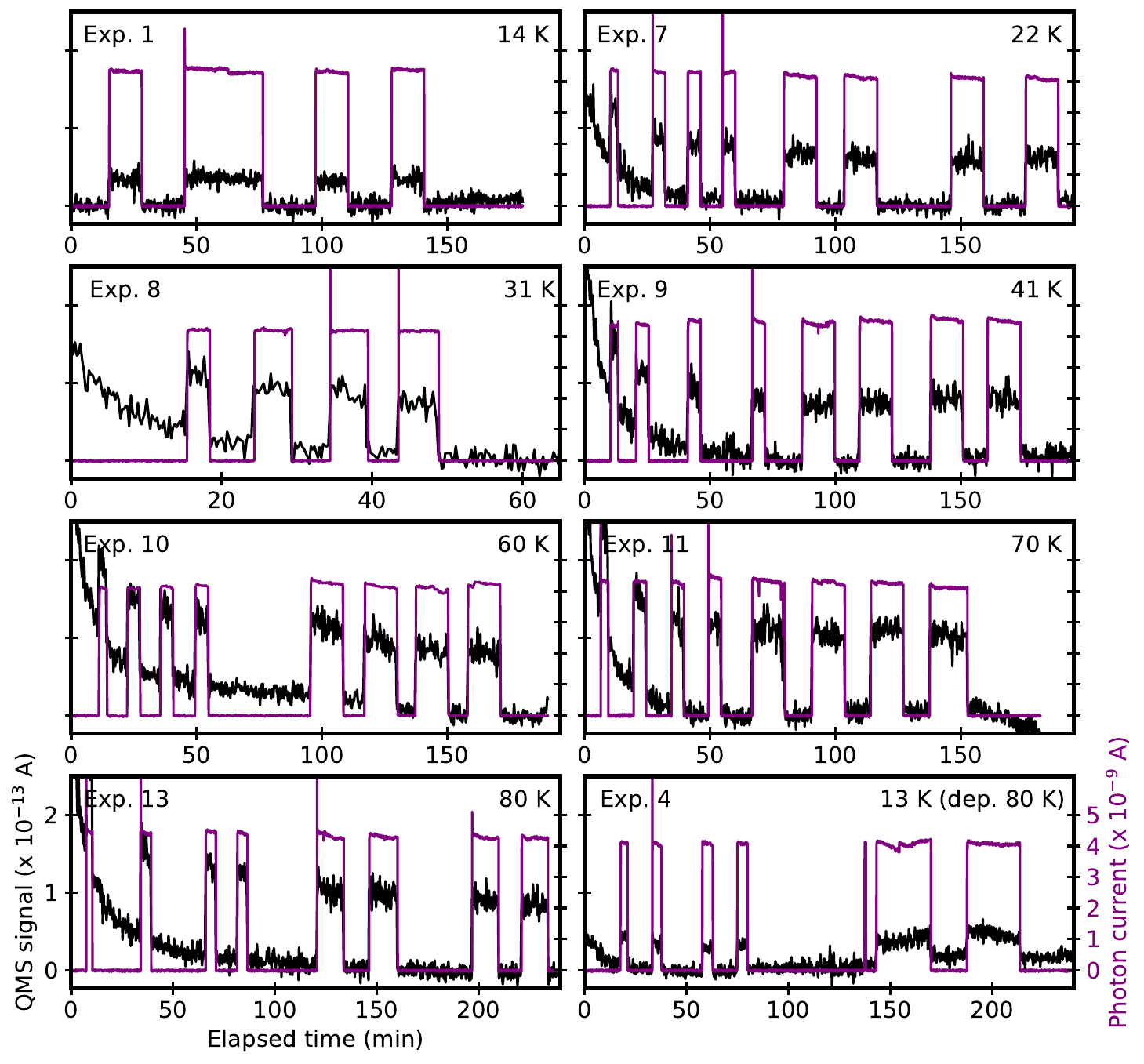}
    \caption{Measured $m/z$ = 64 QMS signal during irradiation of SO$_2$ ice samples at 13$-$80 K (black), along with the irradiated UV flux in every experiment (purple).}
    \label{fig:so2_QMS}
\end{figure*}

\begin{figure*}
    \centering
    \includegraphics[width=0.6\linewidth]{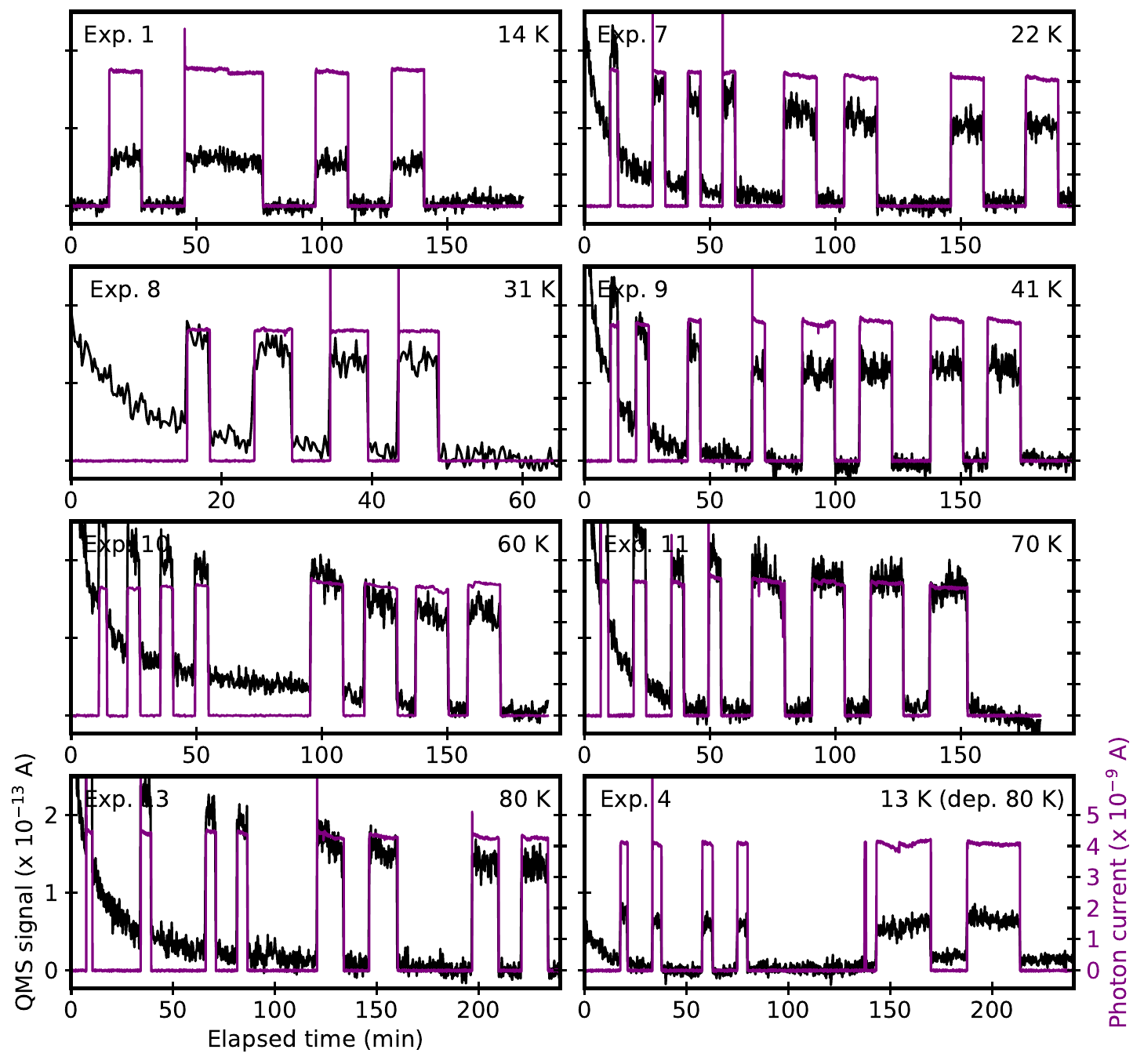}
    \caption{Measured $m/z$ = 48 QMS signal during irradiation of SO$_2$ ice samples at 13$-$80 K (black), along with the irradiated UV flux in every experiment (purple).}
    \label{fig:so_QMS}
\end{figure*}

The $m/z$ = 64  and $m/z$ = 48 QMS signals measured during irradiation of SO$_2$ ice samples at 13$-$80 K are shown in Figures \ref{fig:so2_QMS} and \ref{fig:so_QMS}, respectively.

\section{Modeling the Horsehead PDR}\label{sec:appHH}

\begin{figure}
    \centering
    \includegraphics[width=\linewidth]{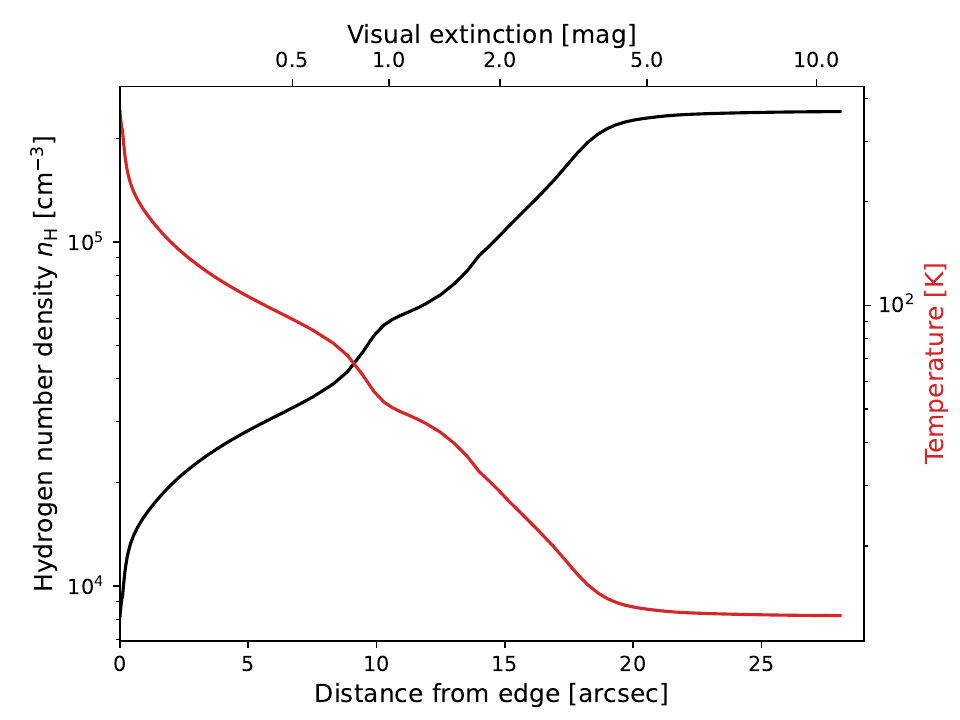}
    \caption{One-dimensional model of the Horsehead PDR. The density and temperature at a given distance from the edge of the PDR are shown as black and gray solid lines, respectively. The visual extinction at each distance from the edge is shown in the upper horizontal axis.}
    \label{fig:hhModel}
\end{figure}

The observed chemical diversity in PDRs poses a challenge for their physical and chemical modeling. 
The physical structure of the Horsehead PDR has been extensively modeled in the literature using static models that assume constant gas density or pressure \citep[see, e.g.,][]{Kaufman1999, LePetit2006}. 
More recently, static isobaric models have been used as an alternative \citep{Bron2018, Joblin2018}. 
The latter scenario is favored by recent observations of molecular tracers performed with ALMA and \emph{Herschel} toward the warm layers of PDRs \citep{Goicoechea2016, Joblin2018, Bron2018, Maillard2021}
We thus adopted the static isobaric approach for our physical modeling of the Horsehead PDR, assuming a FUV field strength of $\chi\sim 110$ in Draine units, equivalent to a FUV field of $G_{0}\sim 186$ in Habing units. The cosmic-ray ionization rate was set to $\zeta_{\rm H_{2}}=4\times 10^{-17}$ s$^{-1}$ \citep{Goicoechea2009}. Using the Meudon PDR code \citep{LePetit2006}, this results in a set of plane-parallel layers shielded from the UV field located at the origin by the visual extinction shown in Fig. \ref{fig:hhModel}. The corresponding density and temperature of each layer are also shown in Fig. \ref{fig:hhModel}.
We note that, even though static isobaric models neglect the pressure gradients observed in PDRs, they remain as good approximations to the physical structure of dynamical photoevaporating PDRs, as shown in \citet{Bron2018}.
In addition, the assumed structure in the static isobaric model does not evolve dynamically. In contrast to our model, UV photons are expected to progressively erode the content of PDRs as the ionization and dissociation fronts move when they overcome the internal pressure of the cloud. In this scenario, photons would constantly interact with fresh, unprocessed gas, potentially increasing the impact of photodesorption on the chemical composition of the PDR. 
According to dynamical models of PDRs, the speed at which the photodissociation front moves is proportional to $G_{0}$/n, being $G_{0}$ the strength of the UV field in terms of the Habing field and $n$ the gas density. In the case of the Orion Bar PDR, this velocity is around 0.2 arcsec in 100 years \citep{Bron2018}. A lower velocity would be expected in the Horsehead PDR due to its lower UV field strength. In Fig. \ref{fig:SO2richIce} the steady state at low extinctions (where the contribution of photodesorption to the SO$_2$ and SO abundances is higher) is reached in $\sim$100 years. As a result, any dynamical effect happening in a shorter timescale would only be observable with a very high angular resolution ($<$0.2 arcsec). Single-dish observations, as the ones used to determine the SO$_2$/SO ratio presented in Fig. \ref{fig:SO2richIce}, would thus be unable to detect chemical changes produced by the moving ionization and dissociation fronts in timescales shorter than 100 years. Therefore, static isobaric models provide a suitable approximation for studying the effect of the SO$_2$ and SO photodesorption yields in the chemistry of the Horsehead PDR.

\begin{figure*}
    \centering
    \includegraphics[width=\linewidth]{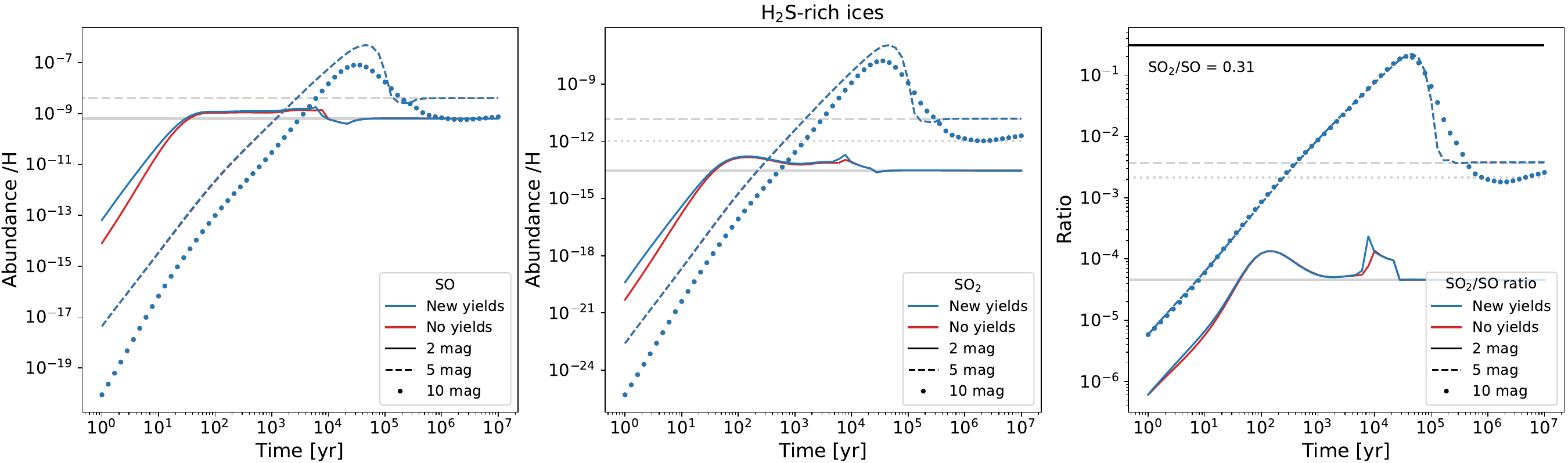}
    \caption{Model predictions of SO (left panel), SO$_{2}$ (middle panel), and the SO$_{2}$/SO ratio (right panel) in a H$_{2}$S-rich ice scenario. The predictions that include our photodesorption yields are plotted in blue, while for the predictions in red the photodesorption yields were set to zero. Predictions at different extinctions are depicted with varying line styles: 2 mag (solid line), 5 mag (dashed line), and 10 mag (dotted line). The steady state values are represented with horizontal gray lines. The SO$_{2}$/SO ratio observed toward the Horsehead PDR is shown as an horizontal black line in the right panel.}
    \label{fig:H2SrichIce}
\end{figure*}

Regarding the chemistry, gas-phase steady-state chemical models have been previously used to explain the chemical makeup of PDRs. However, they are insufficient to reproduce, for instance, the SO$_{2}$/SO ratio observed toward these sources. Therefore, in Sect. \ref{sec:disc_HH} we used instead the time dependent gas-grain chemical model \texttt{Nautilus} \citep{Ruaud2016} for a better characterization of the chemistry in the Horsehead PDR. 
The modeling results for a SO$_2$-rich ice scenario are shown in Fig. \ref{fig:SO2richIce}, whereas Fig. \ref{fig:H2SrichIce} shows the modeling results for a H$_2$-rich ice scenario. 
For these models, the H$_{2}$S photodesorption yield was taken from \cite{fuente17}. 

\bsp	
\label{lastpage}
\end{document}